\documentclass{aa}  

\usepackage{graphicx}
\usepackage{txfonts}
\usepackage{multirow}
\usepackage{caption, threeparttable}
\begin{document}

\newcommand{\teff}{$T_\mathrm{eff}$}
\newcommand{\logg}{$\log g$}
\newcommand{\feh}{[Fe/H]}
\newcommand{\sife}{[Si/Fe]}
\newcommand{\mgfe}{[Mg/Fe]}
\newcommand{\afe}{[$\alpha$/Fe]}
\newcommand{\co}{CO($\nu =7-4$)\,}

\newcommand{\water}{H$_2$O}
\newcommand{\invcm}{cm$^{-1}$}
\newcommand{\kms}{km\,s$^{-1}$}
\newcommand{\mic}{$\mu \mathrm m$}


   \title{The Galactic Chemical Evolution of phosphorus observed with IGRINS}

   \subtitle{}

   \author{G. Nandakumar
            \inst{1}
            \and
          N. Ryde
          \inst{1}
          \and
          M. Montelius
          \inst{2}
          \and
          B. Thorsbro
          \inst{3}
         \and
          H. J\"onsson
          \inst{4}
          \and
          G. Mace
          \inst{5}
          }

   \institute{Lund Observatory, Department of Astronomy and Theoretical Physics, Lund University, Box 43, SE-221 00 Lund, Sweden\\
              \email{govind.nandakumar@astro.lu.se}
    \and 
    Kapteyn Astronomical Institute, University of Groningen, Landleven 12, NL-9747 AD Groningen, the Netherlands
    \and 
    Department of Astronomy, School of Science, The University of Tokyo, 7-3-1 Hongo, Bunkyo-ku, Tokyo 113-0033, Japan
    \and     
    Materials Science and Applied Mathematics, Malmö University, SE-205 06 Malmö, Sweden
    \and
    Department of Astronomy and McDonald Observatory, The University of Texas, Austin, TX 78712, USA}

   \date{Received ; accepted }

 
  \abstract
   {Phosphorus (P) is considered to be one of the key elements for life, making it an important element to look for in the abundance analysis of spectra of stellar systems. Yet, there exists only a handful of spectroscopic studies to estimate the P abundances and investigate its trend across a range of metallicities. This is due to the lack of good P lines in the optical wavelength region and the requirement of careful manual analysis of the blended P lines in near infrared H band spectra obtained with individual observations and surveys like APOGEE.    }
   {Based on a consistent and systematic analysis of high-resolution, near-infrared IGRINS spectra of 38 K giant stars in the Solar neighborhood, we present and investigate the P abundance trend in the metallicity range of -1.2 dex $<$ [Fe/H] $<$ 0.4 dex. Furthermore, we compare this trend with the available chemical evolution models to shed some light on the origin and evolution of P.  }
   {We have observed full H and K band spectra at a spectral resolving power of R=45,000 with IGRINS mounted on the Gemini South telescope, the Discovery Channel Telescope and the Harlan J Smith Telescope at McDonald Observatory. Abundances are determined from spectral lines by modelling the synthetic spectrum that best matches the observed spectrum by $\chi^{2}$ minimisation. For this task we use the Spectroscopy Made Easy (SME) tool in combination with 1D MARCS stellar atmosphere models. The investigated sample of stars have reliable stellar parameters estimated using optical FIES spectra obtained in a previous study of a set of stars called Giants in the Local Disk (GILD). In order to determine the P abundances from the at 16482.92\,\AA\, P line, we need to take special care of the blending the CO($\nu=7-4$) line. With the stellar parameters known, we thus determine the C, N, O abundances from atomic carbon and a range of non-blended molecular lines (CO, CN, OH) which are aplenty in the H band region of K giant stars, assuring an appropriate modelling of the blending CO($\nu=7-4$) line.}
   {We present [P/Fe] vs [Fe/H] trend for K giant  stars in the metallicity range of -1.2 dex $<$ [Fe/H] $<$ 0.4 dex and enhanced P abundances for two metal poor s-rich stars. We find that our trend matches well with the compiled literature sample of prominently dwarf stars and limited number of giant stars. Our trend is found to be higher by $\sim$ 0.05 - 0.1 dex compared to the theoretical chemical evolution trend in \cite{cescutti:2012} resulting from core collapse supernova (type II) of massive stars with the P yields from \cite{Kobayashi:2006} arbitrarily increased by a factor of 2.75 (as suggested by \citealt{cescutti:2012}). Thus the enhancement factor might need to be $\sim$ 0.05 - 0.1 dex higher to match our trend. We also find an empirically determined primary behaviour for phosphorus. Furthermore, the phosphorus abundance is found to be elevated by $\sim$ 0.6 -  0.9 dex in the two s-enriched stars compared to the theoretical chemical evolution trend.  }
   {}


   \maketitle
%

\section{Introduction}
\label{sec:intro}

Analogous to fossils here on the Earth, 
light from stars can be used to understand the formation and evolution history of stellar populations. This is possible through detailed spectroscopic analysis of stellar spectra from which chemical compositions of each star and hence their parent molecular clouds can be determined. Individual elemental abundances derived from stellar spectra in conjunction with chemical evolution models have also been used to infer their nucleosynthetic origin, trace possible star progenitors, predict their stellar yields, and impose constraints on their formation inside stars. For example, $\alpha$ elements like O, Mg, Si, Ca, and Ti are formed in massive stars during hydrostatic and explosive burning phases, and are released to the interstellar medium (ISM) when they undergo type II supernova (SN) explosion \citep{Chieffi:2004,Nomoto:2013,Matteucci:2021}.

There are many elements whose origin and formation sites are still in debate or unknown. This is due to various reasons like absence of measurable lines in commonly observed wavelength ranges, unreliable or unmeasured atomic data, weak/strong lines that are difficult to model etc. Until a decade ago, phosphorus was one such element that remained unexplored in long lived stars of spectral type F, G, or K owing to the 
absence of phosphorus lines
in the commonly observed wavelength ranges of these stars \citep{Struve:1930,Caffau:2011}. Phosphorus is specifically interesting since it is among the key elements for biology on Earth: H, C, N, O, P and S. Assuming a 1:1 correlation between elemental abundance in the star and the material out of which planets form, phosphorus detection in exoplanet hosts could prove useful in the search for life \citep{Hinkel:2020}. Thus, it is important to identify and model measurable phosphorus lines for which reliable phosphorus abundance can be determined for large stellar samples. It will then be possible to investigate the origin and evolution of phosphorus based on the comparison of its observed trend versus metallicity and other elements to those predicted by chemical evolution models. 

One of the first attempt to investigate phosphorus trend with metallicity was by \cite{Caffau:2011} with P abundances measured from the infrared weak P I lines of the multiplet 1, at 1050-1082 nm in the CRIRES spectra for 20 cool dwarf stars in the Galactic disk. They showed a decreasing trend of [P/Fe] with increasing metallicity in the range of -1.0 $<$ [Fe/H] $<$ 0.3 dex which is different from the trend they found for other light odd-Z elements, sodium and aluminium. These abundance measurements were used by \cite{cescutti:2012} to construct and compare chemical evolution models with phosphorus and metallicity yields from \cite{Woosley:1995} and \cite{Kobayashi:2006} and found that yields from massive stars have to be increased by a factor of 2.75 to match the observational trend. \cite{Jacobson:2014} and \cite{Roederer:2014} extended the lower metallicity range of the [P/Fe] vs [Fe/H] trend down to -3.8 dex determining an upper limit value for [P/Fe] with a P I doublet in the ultraviolet, 2135.465\,\AA\, and 2136.182\,\AA\, using the Hubble Space Telescope–Space Telescope Imaging Spectrograph (STIS; \citealt{Kimble:1998, Woodgate:1998}). Using near-infrared (NIR) high resolution spectrograph GIANO \citep{Origlia:2014} at Telescopio Nazionale Galileo (TNG), and with the same lines used in \cite{Caffau:2011}, \cite{Caffau:2016} and \cite{Caffau:2019} found a consistent phosphorus trend with metallicity using $\sim$43 stars. Similarly, \cite{Maas:2017} and \cite{Maas:2019} estimated P abundances for FGK dwarfs, giants and Hyades open cluster members using the aforementioned P I lines based on observations with Phoenix high-resolution spectrometer \citep{Hinkle:1998} on the Kitt Peak National Observatory Mayall 4 m telescope and the Gemini South telescope. They also found consistent [P/Fe] vs [Fe/H] trends with the literature and estimated solar values for abundance ratios of phosphorus with elements like O, Mg, Si and Ti pointing towards a similar formation site for phosphorus as these elements. Using spectra observed with the Habitable Zone Planet Finder (HPF) on the 10 m Hobby–Eberly Telescope (HET) at McDonald Observatory \citep{Mahadevan:2012,Mahadevan:2014}, \cite{Sneden:2021} estimated P abundances from the same set of P I lines for 13 field red horizontal branch (RHB) stars. Very recently, \cite{Maas:2022} used spectra from the same instrument to estimate P abundances for a sample of 163 FGK dwarfs and giants and find thick disk stars to have approximately 0.1 dex higher [P/Fe] than thin disk stars.

Among the large scale spectroscopic surveys, Apache Point Observatory Galactic Evolution Experiment/APOGEE \citep{Majewski:2017} which observe in the near infrared H-band (15,100--17,000\,\AA, R$\sim$22,500) regime, have identified two P I lines, 15711.1 and 16482.9\,\AA\ \citep{Shetrone:2015}, but the industrial pipeline used by APOGEE to analyse the spectra is not optimal for weak and blended lines like these. Hence, APOGEE DR12 \cite{Holtzman:2015} does not include P abundance measurements. However, \cite{Hawkins:2016} re-analysed APOGEE spectra for APOGEE$+$Kepler stellar sample (APOKASC) of $\sim$2000 giants with the BACCHUS code \citep{Masseron:2016} and estimated P abundances from the aforementioned two lines, finding decreasing [P/Fe] trend with increasing metallicity. The recent data releases, APOGEE DR13 - DR16 (e.g. \citealt{jonsson:2018}), includes P though it is cautioned to be one of the most uncertain element in APOGEE. As a result, the latest data release of APOGEE (DR17, Holtzman et al. in prep.), decided not to include any P abundances. But \cite{Hayes:2022} has done a re-analysis of APOGEE data with BACCHUS to measure P, among other difficult-to-measure elements in APOGEE.

 A similar type of re-analysis of APOGEE spectra led to the discovery of 15 P-rich stars by \cite{Masseron:2020}. The elemental abundance patterns of these stars with enhanced Mg, Si, Al, and s-process elements have challenged theoretical models, suggesting these over abundances could have formed from a new s-process site \citep{Masseron:2020b}. The same two P lines in higher resolution  (R$\sim$45,000) IGRINS spectra have been used by \cite{Afsar:2018}, \cite{Topcu:2019} and \cite{Topcu:2020} to estimate [P/Fe] for three field red horizontal branch stars, approximately 10 giant stars in the open clusters NGC 6940 and NGC 752 respectively.

The above mentioned studies indicate a dearth of reliable phosphorus abundance estimates, especially for giant stars. Even though giants are less interesting in the context of searching for habitable planets, they allow us to probe more distant populations. Thus there is an evident necessity for more high resolution spectroscopic observations as well as systematic analysis to increase the sample size of stars with reliable P estimates especially for giant stars. Here, we present a consistent, detailed and systematic analysis of the phosphorus line at 16482.92\,\AA\, in the high resolution near infrared H band IGRINS spectra of 38 K giant stars in the solar neighborhood to determine their phosphorus abundances. Our sample of stars are a subset of the Giants in the Local Disk (GILD) sample (in the same way as \cite{Ryde:2020} and \cite{montelius:22}) and have reliable stellar parameters determined from optical FIES spectra. In section~\ref{sec:obs}, we briefly describe the IGRINS observations and data reduction carried out to obtain the final spectra. The determination of phosphorus from spectra using SME is described in the section~\ref{sec:analysis}, followed by results and discussion in sections~\ref{sec:results} and \ref{sec:discussion} respectively. Finally, we make the concluding remarks in the section~\ref{sec:conclusion}.

 \begin{figure}
  \includegraphics[width=\columnwidth]{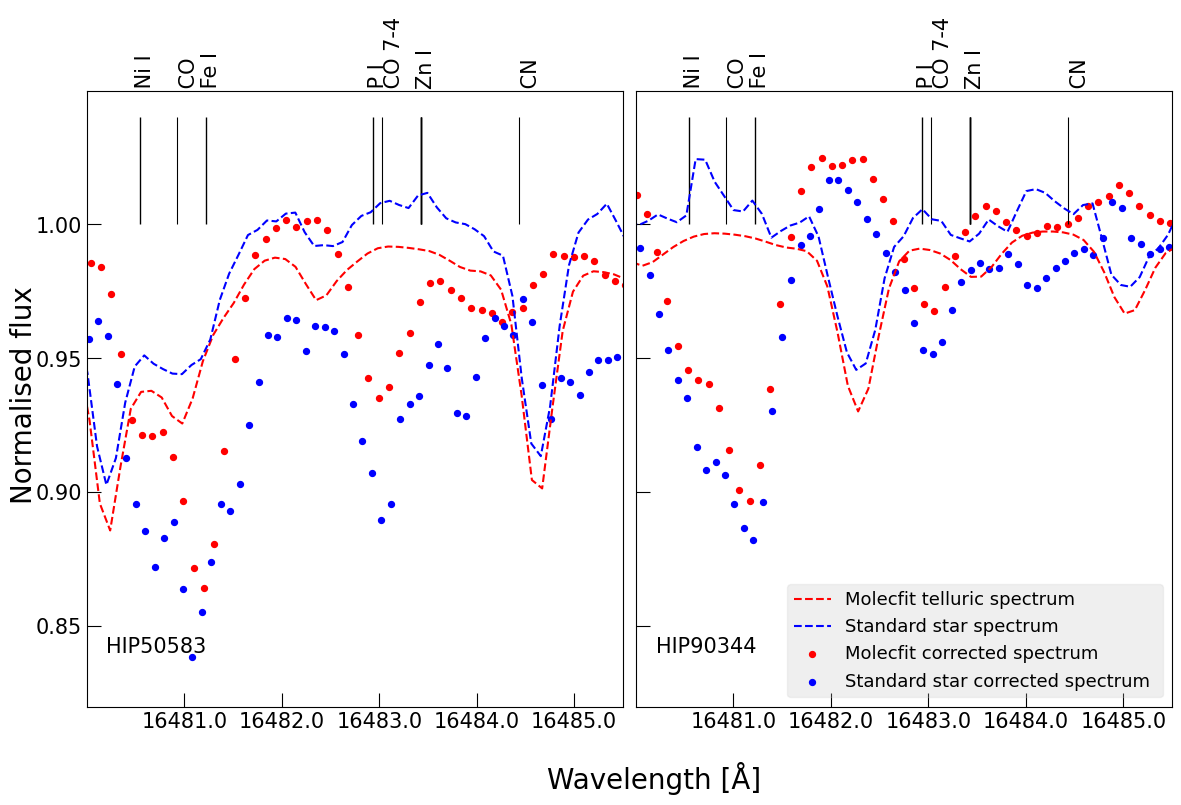}
  \caption{The Molecfit corrected spectrum (red dots) and the standard star corrected spectrum (blue dots) around the P I line at 16482.92\,\AA\ for two stars in our sample. The normalised telluric spectra from the standard star observation and from using the Molecfit software tool are shown in blue dashed line and red dashed line, respectively. Since the telluric standard star spectra show unexpected spurious features at the wavelengths of the P line, we eliminate the telluric lines with Molecfit.
  Using the standard star spectra with the spurious emission feature would lead to an artificial increase in the strength of the P I line.}
  \label{fig:molecfit}%
\end{figure}

\begin{figure*}
  \includegraphics[width=\textwidth]{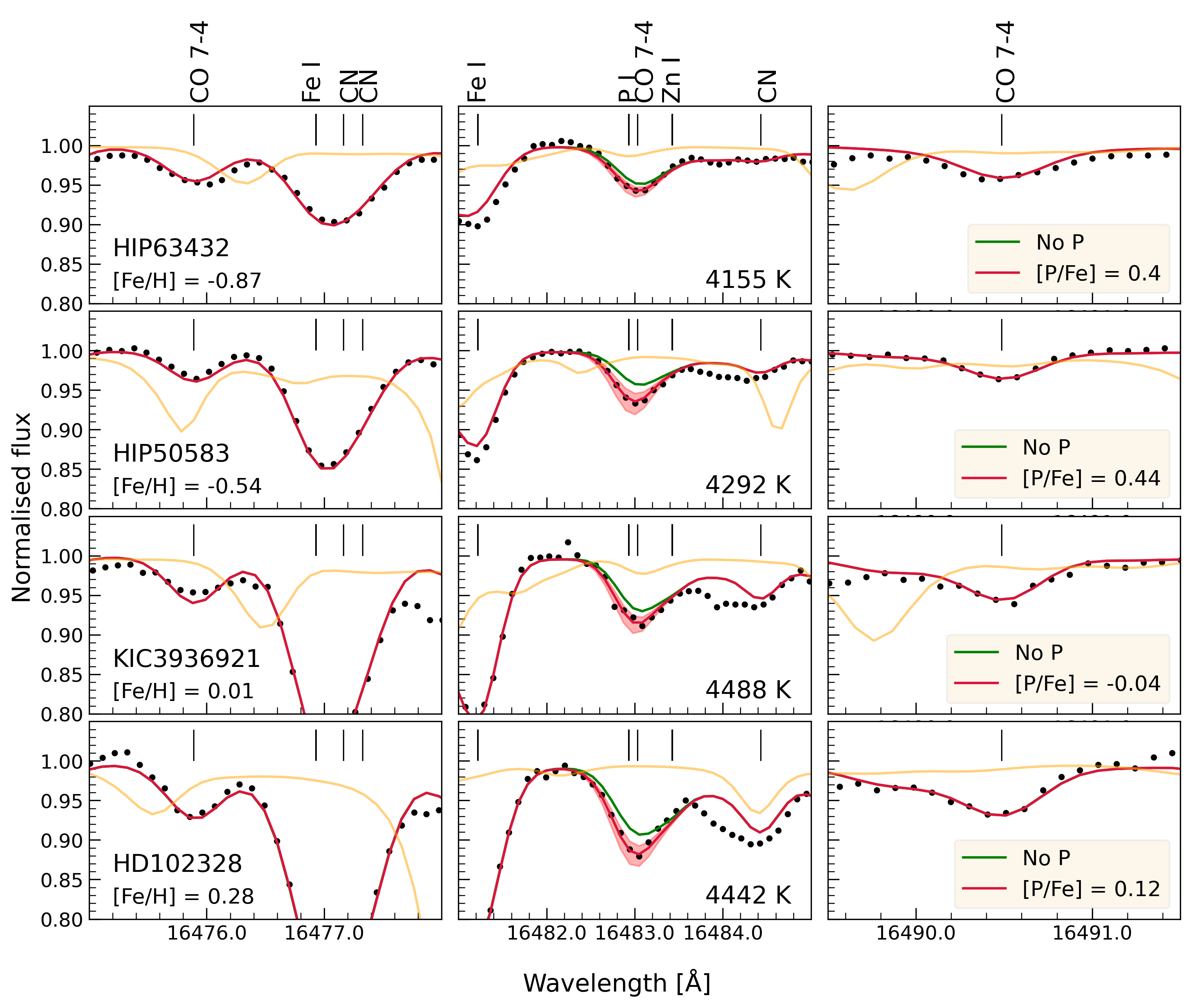}
  \caption{Fits (red line) to the observed spectra (black points) for 4 stars chosen to represent typical spectrum at metallicities of -0.87 dex (HIP63432), -0.54 dex (HIP50583), 0.01 dex (KIC3936921) and 0.28 dex (HD102328). Each row represent one star with stars arranged in the order of increasing metallcity from top to bottom. The segment of the corresponding spectrum with the P line at 16482.92\,\AA\,is shown in the middle panel in each row. In the left and right panels, spectra of the neighbouring \co lines are plotted. In each panel, the green line shows the synthetic spectrum without P or only showing the CO($\nu =7-4$) blend. The telluric spectrum used in the reduction is shown in yellow in order to indicate the pixels affected by telluric correction. The red band in the middle panel for the P line shows the variation of synthetic spectrum with $\Delta$[P/Fe] = $\pm$0.3 dex variation. The effective temperature, \teff, of each star is listed in the middle panels. }
  \label{fig:selspectra}%
\end{figure*}

\begin{figure}
  \includegraphics[width=\columnwidth]{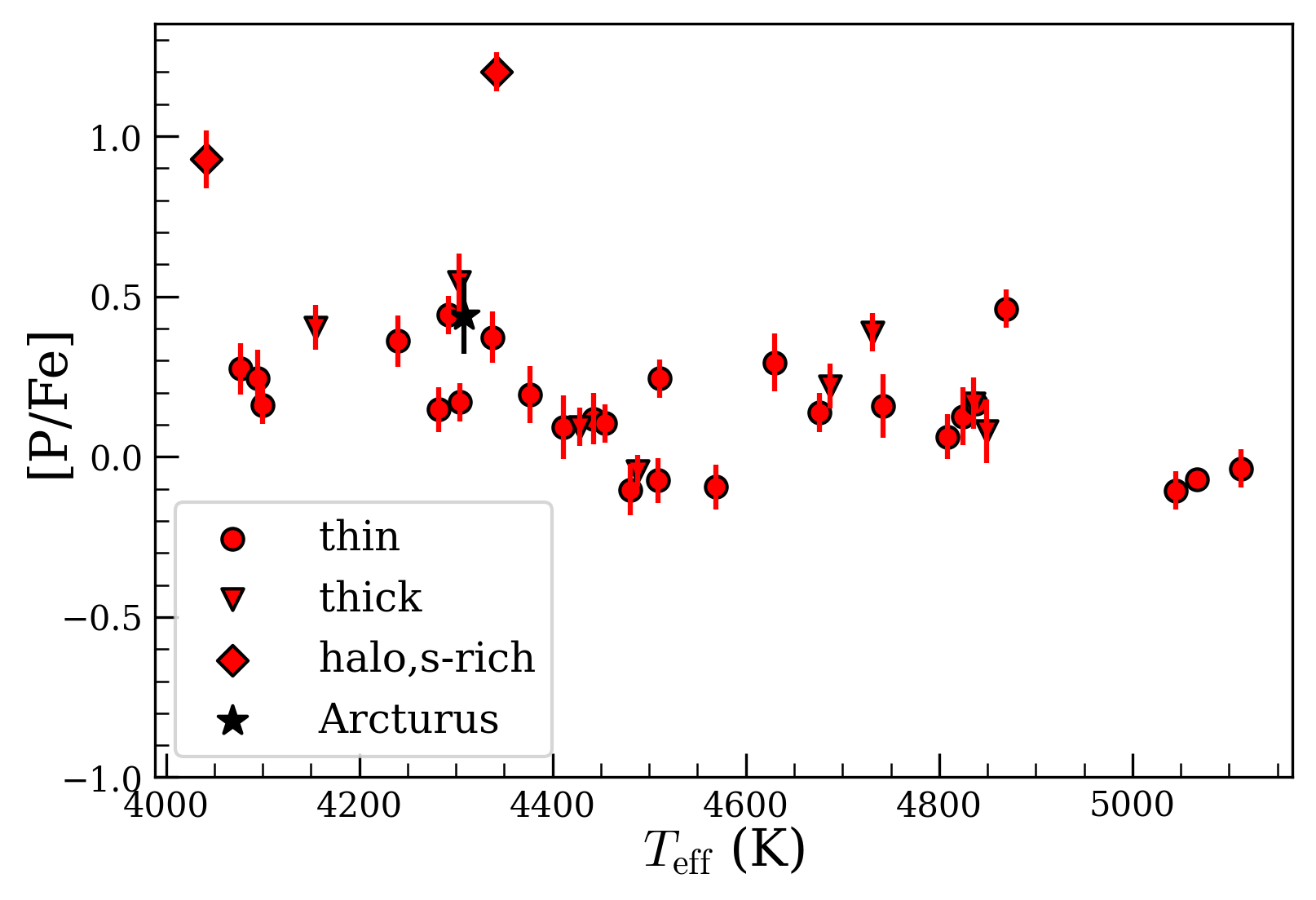}
  \caption{[P/Fe] vs [\teff] for stars in our sample. Different symbols represent stellar populations to which each star belongs.}
  \label{fig:pteff}%
\end{figure}

\begin{table*}
\caption{ Observational details of K giant stars. }\label{table:obs}
\centering
\begin{tabular}{c c c c c c c}
\hline
\hline
 Star  & 2MASS ID  & H$_{2MASS}$  & K$_{2MASS}$  &  Civil  & Telescope  & Exposure  \\

   &  & (mag) & (mag) & Date &  & s  \\
\hline
 $\mu$leo & J09524585+2600248  &    1.3 &  1.2 &  2016 Jan. 31  & HJST  & 26 \\
 $\epsilon$Vir  &   J13021059+1057329 & 0.9 &  0.8 & 2016 Feb. 2  & HJST  & 26  \\
 $\beta$Gem &   J07451891+2801340 & -1.0 & -1.1 & 2016 Jan. 30  & HJST  & 23  \\
 HD102328  &  J11465561+5537416 &   2.9 &  2.6 & 2016 Feb. 2  & HJST  & 32 \\
 HIP50583  &   J10195836+1950290 & -0.8 & -0.8 & 2016 June 20 & HJST  & 29  \\ 
 HIP63432  &   J12595500+6635502 & 2.4 &  2.1 & 2016 May 29  & HJST  & 180 \\ 
 HIP72012 &  J14434444+4027333 & 2.6 &  2.4 & 2016 June 16  & HJST  & 120 \\ 
 HIP90344  &  J18255915+6533486 & 2.2 &  2.1 & 2016 June 15 & HJST  & 120   \\ 
 HIP96014 &   J19311935+5018240 & 2.9 &  2.5 & 2016 June 15  & HJST  & 120 \\ 
 HIP102488  &  J20461266+3358128 & 0.2 &  0.1 & 2016 June 19  & HJST  & 26  \\ 
2M16113361  &  J16113361+2436523  &  7.9  &  7.7   &  2016 July 26  &  HJST  &  180 \\
2M17215666  &   J17215666+4301408 &   7.6 &  7.5 & 2016 July 25 & HJST  & 1080   \\ 
KIC3748585  &   J19272877+3848096 &  6.4 &  6.3 & 2016 Nov. 17& DCT & 240  \\ 
KIC3936921  &  J19023934+3905592  &  8.3  &  8.1   &  2016 Nov 23  &  DCT  &  120 \\
 KIC3955590 &  J19272677+3900456 & 7.8 &  7.7 & 2016 Nov. 23  & DCT   & 480 \\ 
 KIC4659706  & J19324055+3946338 &    7.6 &  7.4 & 2016 Nov. 19 & DCT & 480  \\ 
KIC5113061  &  J19413439+4017482  &  8.2  &  8.0   &  2016 Nov 22  &  DCT  &  90 \\
 KIC5113910  &   J19421943+4016074 & 8.2 &  8.0 & 2016 Nov. 22  & DCT   & 720 \\ 
 KIC5709564  &  J19321853+4058217 &  7.6 &  7.5 & 2016 Nov. 23  & DCT   & 480  \\
 KIC5779724  &   J19123427+4105257 &  8.0 &  7.8 & 2016 Dec. 09  & DCT  & 600   \\ 
 KIC5859492  &  J19021718+4107236 &  7.9 &  7.8 & 2016 Nov. 23  & DCT   & 480 \\
 KIC5900096  &  J19515137+4106378 & 6.0 &  5.8 & 2016 Nov. 22  & DCT  & 120 \\ 
 KIC6465075 & J19512404+4149284  &   8.0 &  7.9 & 2016 Nov. 23  & DCT  & 480 \\
 KIC6837256 &  J18464309+4223144 &   8.8 &  8.7 & 2016 Nov. 23  & DCT   & 720 \\
KIC7006979  &  J18433894+4235295  &  7.5  &  7.6   &  2016 Nov 18  &  DCT  &  90 \\
 KIC10186608 & J18444919+4717434 &     8.9 &  8.8 & 2016 Nov. 17  & DCT   & 960 \\ 
 KIC11045542  &  J19530590+4833180 &  8.4 &  8.2 & 2016 Dec. 11 & DCT & 2000  \\ 
KIC11342694 &  J19110062+4906529 &  7.6 &  7.4 & 2016 Nov. 17 & DCT & 480  \\ 
KIC11444313  &   J19014380+4923062 &   9.1 &  9.0 & 2016 Nov. 23  & DCT   & 720\\
 KIC11569659 &  J19464387+4934210 &  9.4 &  9.3 & 2016 Dec. 9 & DCT & 2400 \\
ksiHya  & J11330013-3151273 & 1.5  & 1.4  &  2016 June 30 & HJST & 8 \\
HD175541  & J18554089+0415551 & 5.9  & 5.8  &  2022 May 09 & Gemini South & 80 \\
HD176981  & J19022156+0822248  & 2.6  & 2.5  &  2022 May 24 & Gemini South & 8 \\
HD206610  & J21432490-0724296 & 6.1  & 5.9  &  2022 May 24 & Gemini South & 80 \\
HD76445  & J08564599+1728530 & 5.4  & 5.3  &  2022 Jan. 24 & Gemini South & 20 \\
NGC6705\,1184  & J18513901-0613070 & 10.6  & 10.4  &  2022 May 09 & Gemini South & 160 \\
NGC6705\,1423  & J18505581-0618147 & 7.9  & 7.7  &  2022 May 24 & Gemini South & 200 \\

\hline
\hline                                 
\end{tabular}
\tablefoot{\\DCT: the Discovery Channel Telescope, a 4.3 m telescope at Lowell Observatory, Arizona.\\
HJST: the  Harlan J Smith Telescope, a 2.7 m telescope at McDonald Observatory, Texas. \\
Gemini South : 8.1-meter Gemini South telescope.}
\end{table*}

\section{Observations and Data reduction}
\label{sec:obs}

In this work, we have carried out near infrared spectroscopic observations of 38 K giant stars with Immersion GRating INfrared Spectrograph \citep[IGRINS;][]{Yuk:2010,Wang:2010,Gully:2012,Moon:2012,Park:2014,Jeong:2014} in 2016 on the 4.3-meter Discovery Channel Telescope (DCT; now called the Lowell Discovery Telescope) at Lowell Observatory \citep{Mace:2018}, and on the 2.7-meter Harlan J. Smith Telescope at McDonald Observatory \citep{Mace:2016}, as well as on Gemini South telescope \citep{Mace:2018} in early 2022 
as part of the poor-weather programs GS-2021B-Q416 and GS-2022A-Q408. IGRINS provides spectra spanning the full H and K bands (1.45 - 2.5 $\mu$m) with a spectral resolving power of R $\sim$ 45000. Details of observations are listed in the Table~\ref{table:obs}.

We selected the stars from among nearly 500 local disk giants from the observing program with Fiber-fed Echelle Spectrograph \citep[FIES;][]{Telting:2014} at the Nordic Optical Telescope (NOT) started in \cite{jonsson:17}. The stellar parameters for these stars have been determined through a careful optical spectroscopic analysis using the same tools for the spectral analysis as presented here (J\"onsson et al. in prep). We also use the high resolution (R $\sim$ 100,000) infrared spectrum of Arcturus from the Arcturus atlas \citep{Hinkle:1995}.

The IGRINS observations were carried out in an ABBA nod sequence along the slit permitting sky background subtraction. With the aim of achieving signal-to-noise ratios (SNR) of at least 100, exposure times were set to $\sim$ 10 to 2500 seconds. Telluric standard stars were chosen to be rapidly rotating, late B to early A dwarfs and observed close in time and at a similar air mass as the science targets. We used the IGRINS PipeLine Package \citep[IGRINS PLP;][]{Lee:2017} to optimally extract the telluric corrected,  wavelength calibrated spectra after flat-field correction and A-B frame subtraction. The spectral orders of the science targets and the telluric standards are subsequently stitched together after normalizing every order and then combining them in {\tt iraf} \citep{IRAF}, excluding the low signal-to-noise edges of every order. This resulted in one normalized stitched spectrum for the entire H band. However, to take care of any modulations in the continuum levels of the spectra we put large attention in defining specific local continua around the spectral line being studied. This turns out to be an important measure for accurate determinations of abundances, see for example \citet{pablo:20}.

The standard procedure for eliminating the contaminating telluric lines is to divide with a telluric standard-star spectrum, showing only telluric lines and mostly no stellar features. This works very well for most wavelength regions. Apart from some broad Brackett lines of hydrogen, some spurious broadband spectral features might, however, turn up in the telluric standard-star spectrum.  For many of our observed stars, the corresponding telluric standard-star spectra 
indeed show broadband (ranging over more than 5\,\AA, i.e. over several resolution elements) and non-flat features affecting the normalized science spectra just in the wavelength region of the P line. There are no such features expected in the telluric spectrum in this wavelength region, where only CH$_{4}$ absorption lines are expected. These observed spurious problems with the telluric standard-star spectra therefore make a proper continuum normalisation of the science spectra impossible. 
In order to rectify this problem we, therefore, carried out telluric correction using the Molecfit software tool \citep{Smette:2015,Kausch:2015} for the spectra surrounding the P line. This tool models and fits the telluric lines in the observed science spectrum for various telluric molecules (H$_{2}$O, CO$_{2}$, CH$_{4}$, O$_{2}$ etc.). Molecfit requires the temperature, pressure, and humidity during the time of the target observation as inputs in order to model the Earth's atmosphere using a state-of-the-art radiative transfer code. The Molecfit GUI tool also provides the option to choose the wavelength regions to be included (strong telluric lines) and thus define masks which are then used to identify the lines to be fitted. Finally, the model parameters obtained by fitting the chosen lines (masks) are used to model the telluric spectrum in the full wavelength range with which the science spectrum is corrected. Figure~\ref{fig:molecfit} shows the difference between the molecfit corrected spectrum (red dots) and the standard star corrected spectra (blue dots) for two stars in our sample. The telluric spectra from the standard star observations (blue dashed lines) with problematic continuum normalisation result in artificially stronger lines, especially for our line of interest at 16481.92\,\AA. Meanwhile, the telluric spectra from Molecfit (red dashed lines) show a reliable continuum normalization and thus provide telluric corrected science spectra from which the phosphorus abundance can be determined.

In all our spectral plots we also show the corresponding telluric spectra from Molecfit (in yellow) which is used to eliminate the telluric lines. We do this in order to show how good the elimination procedure actually works. Furthermore, it also shows where the telluric lines originally lay and where the signal-to-noise, therefore, is expected to be lower, as well as where spectral residuals might sometimes be identified, especially from the strongest telluric features.

Finally, for the wavelength solutions, sky OH emission lines are used \citep{Han:2012,Oh:2014} and the spectra are subsequently shifted to laboratory wavelengths in air after a stellar radial velocity correction. In addition, we make sure to carefully eliminate obvious cosmic-ray signatures in the spectra.


\section{Analysis}
\label{sec:analysis}

\subsection{Phosphorus lines in IGRINS}
\label{sec:Plines}


There are two phosphorus lines in the H band spectra that have been used in the literature: the lines at 15711.52\,\AA\, and 16482.92\,\AA. Upon initial checks, the line at 15711.52\,\AA\, in our spectra is found to be severely affected by telluric lines making P abundance determination with this line impossible. Hence we only use the line at 16482.92\,\AA\ in our analysis. For this line, there is a theoretically determined log($gf$)=-0.27 \citep{Biemont:1994}, but when using this value, the line is synthesized too strong when comparing to a solar spectrum. For this reason, \cite{Afsar:2018a} determined astrophysical log($gf$) value of -0.56 for this line by applying reverse solar analysis using the high resolution infrared solar flux spectrum of \cite{Wallace:2003} and a solar P abundance value of 5.41 from \cite{Asplund:2009}. Similarly we determined an astrophysical log($gf$) value of -0.51 based on reverse solar analysis with the same solar flux spectrum. The difference of 0.05  with respect to the value from \cite{Afsar:2018a} is due to the lower solar P abundance value of 5.36 we adopted from \cite{solar:sme}.

Due to the odd number of protons in the phosphorus nucleus, the spectral lines are likely to be affected by hyperfine structure splitting (see e.g. \cite{Pendlebury:1964}). The line we measure our abundances from does not have information on the strength of the hyperfine structure splitting, but as the line is weak we do not expect it to impact our results significantly.


We also determined astrophysical log($gf$) values based on reverse solar analysis for the spectral lines of light odd-Z (Na, Al, and K) and even-Z elements (Mg, Si, and S) used in this work (see Section~\ref{sec:elements} and Table~\ref{table:lines} below).
 For the  molecular lines, we used the line data for CO from \citet{li:2015}, for CN from \citet{sneden:2014} and for OH from from \citet{brooke:2016}. 
 In the section below, we explain in detail the determination of P abundance from this line.

\subsection{Determination of P abundance}
\label{sec:Pabund}

We determine P abundance from the observed spectra using the Spectroscopy Made Easy code \citep[SME;][]{sme,sme_code} that generates synthetic spectra for a set of stellar parameters by calculating the radiative transfer and interpolating in a grid of hydrostatic one-dimensional (1D) MARCS stellar atmosphere models \citep{marcs:08} in spherical geometry assuming LTE, chemical equilibrium, homogeneity, and conservation of the total flux. SME generates and fits multiple synthetic spectra for the chosen line of interest by varying its elemental abundance (set as a free parameter). The final abundance of the line is chosen to be the one that corresponds to the synthetic spectrum that best matches the observed spectrum by means of $\chi^{2}$ minimisation method.   

As mentioned in the Section~\ref{sec:obs}, the stars analysed in this work are a subset of the GILD stellar catalogue (J\"onsson et al. in prep., which builds upon and improves the analysis described in \citet{jonsson:17}). These stars were also used in, e.g., \citet{jonsson:17,forsberg:19,Ryde:2020,montelius:22}, and Forsberg et al., in press). The fundamental stellar parameters, namely effective temperature (\teff), surface gravity (\logg), metallicity (\feh), and  microturbulence ($\xi_\mathrm{micro}$) are estimated from fitting synthetic spectra for unsaturated and unblended Fe I and Fe II lines, Ca I lines, and \logg\, sensitive Ca I line wings, while \teff\,, \logg\,, \feh\,, $\xi_\mathrm{micro}$, and [Ca/Fe] are set as free parameters in SME. They have been benchmarked against independently determined effective temperatures, \teff, from angular diameter measurements and surface gravities, \logg, from asteroseismological measurements \citep[see][for more details]{jonsson:17}. With uncertainties of $\pm$50K for \teff, $\pm$0.15 dex for \logg, $\pm$0.05 dex for \feh, and $\pm$0.1 km/s for $\xi_\mathrm{micro}$, we have precise stellar parameters.

\begin{table*}
\caption{ Stellar parameters, stellar population from J\"onsson et al. in prep. and $v_{\rm macro}$], [P/Fe], $\sigma$[P/Fe] and signal-to-noise ratio (SNR) determined in this work. }\label{table:all}
\begin{tabular}{c c c c c c c c c c}
\hline
\hline
 Star  & Population  & T$_\mathrm{eff}$ & $\log g$  & [Fe/H]  &  $v_{\rm micro}$  &  $v_{\rm macro}$ & [P/Fe] & $\sigma$[P/Fe] & SNR  \\
 \hline
  & & K & log(cm/s$^{2}$) & dex & Km/s & Km/s & dex & dex & per pixel \\
\hline
$\alpha$Boo$^{*}$   &  thick  &  4308  &  1.7  &  -0.55  &  1.8  &  4.9  &  0.44  &  0.12 & 297 \\
$\mu$Leo  &  thin  &  4494  &  2.5  &  0.27  &  1.5  &  6.7  &  0.29  &  0.08 & 169 \\ 
$\epsilon$Vir  &  thin  &  5112  &  3.0  &  0.11  &  1.6  &  7.7  &  -0.04  &  0.06 & 300\\ 
betaGem  &  thin  &  4809  &  2.8  &  -0.02  &  1.4  &  6.9  &  0.06  &  0.07 & 336\\ 
HD102328  &  thin  &  4442  &  2.5  &  0.28  &  1.5  &  7.1  &  0.12  &  0.08 & 250\\ 
HIP50583  &  thin  &  4292  &  1.6  &  -0.54  &  1.6  &  5.6  &  0.44  &  0.06 & 86\\ 
HIP63432  &  thick  &  4155  &  1.3  &  -0.87  &  1.9  &  7.5  &  0.4  &  0.07 &  275 \\ 
HIP72012  &  thin  &  4077  &  1.4  &  -0.28  &  1.5  &  6.3  &  0.28  &  0.08 &  285 \\ 
HIP90344  &  thin  &  4454  &  2.2  &  -0.39  &  1.4  &  7.0  &  0.1  &  0.06 &  282 \\ 
HIP96014  &  thin  &  4240  &  1.6  &  -0.43  &  1.7  &  7.8  &  0.36  &  0.08 &  414 \\ 
HIP102488  &  thin  &  4742  &  2.5  &  -0.21  &  1.6  &  6.0  &  0.16  &  0.1 &  343 \\ 
2M16113361  &  halo,s-rich  &  4042  &  1.2  &  -1.06  &  1.9  &  6.4  &  0.93  &  0.09 &  160 \\ 
2M17215666  &  halo,s-rich  &  4342  &  1.6  &  -1.11  &  1.6  &  6.7  &  1.2  &  0.06 &  165 \\ 
KIC3748585  &  thin  &  4569  &  2.6  &  0.03  &  1.3  &  5.6  &  -0.09  &  0.07 &  205 \\
KIC3936921  &  thick  &  4488  &  2.2  &  0.01  &  1.6  &  5.4  &  -0.04  &  0.05 &  130 \\
KIC3955590  &  thin  &  4411  &  2.2  &  0.03  &  1.6  &  5.7  &  0.09  &  0.1 &  140 \\ 
KIC4659706  &  thick  &  4428  &  2.5  &  0.24  &  1.4  &  5.4  &  0.09  &  0.06 &  86 \\ 
KIC5113061  &  thin  &  4100  &  1.6  &  -0.09  &  1.8  &  5.5  &  0.16  &  0.06 &  164 \\ 
KIC5113910  &  thin  &  4338  &  1.7  &  -0.48  &  1.6  &  6.2  &  0.37  &  0.08 &  182 \\ 
KIC5709564  &  thick  &  4687  &  2.2  &  -0.35  &  1.7  &  7.0  &  0.22  &  0.07 &  137 \\ 
KIC5779724  &  thick  &  4303  &  1.6  &  -0.45  &  1.6  &  5.7  &  0.54  &  0.09 &  215 \\ 
KIC5859492  &  thin  &  4511  &  2.4  &  0.09  &  1.5  &  5.5  &  0.24  &  0.06 &  180 \\ 
KIC5900096  &  thin  &  4480  &  2.5  &  0.23  &  1.5  &  5.6  &  -0.1  &  0.08 &  179 \\ 
KIC6465075  &  thin  &  4825  &  2.8  &  -0.37  &  1.3  &  5.5  &  0.13  &  0.09 &  146 \\ 
KIC6837256  &  thick  &  4731  &  2.3  &  -0.73  &  1.5  &  5.6  &  0.39  &  0.06 &  148 \\ 
KIC7006979  &  thin  &  4869  &  2.6  &  -0.34  &  1.6  &  5.3  &  0.46  &  0.06 &  104 \\ 
KIC10186608  &  thin  &  4676  &  2.5  &  -0.12  &  1.5  &  5.8  &  0.14  &  0.06 &  91 \\ 
KIC11045542  &  thin  &  4304  &  1.6  &  -0.65  &  1.5  &  5.9  &  0.17  &  0.06 &  294 \\ 
KIC11342694  &  thin  &  4509  &  2.8  &  0.14  &  1.3  &  5.3  &  -0.07  &  0.07 &  139 \\ 
KIC11444313  &  thin  &  4630  &  2.3  &  -0.18  &  1.5  &  5.3  &  0.29  &  0.09 &  160 \\ 
KIC11569659  &  thin  &  4838  &  2.4  &  -0.43  &  1.6  &  5.8  &  0.16  &  0.03 &  246 \\
$\xi$Hya  &  thin  &  5045  &  2.8  &  0.31  &  0.5  &  8.1  &  -0.1  &  0.06 &  176 \\ 
HD175541  &  thin  &  5067  &  3.5  &  -0.15  &  1.2  &  3.7  &  -0.07  &  0.02 &  115 \\ 
HD176981  &  thin  &  4095  &  1.6  &  -0.22  &  1.8  &  5.7  &  0.24  &  0.09 &  95 \\ 
HD206610  &  thick  &  4849  &  3.3  &  0.04  &  1.2  &  3.2  &  0.08  &  0.1 &  152 \\ 
HD76445  &  thick  &  4836  &  3.1  &  -0.22  &  1.4  &  3.6  &  0.17  &  0.08 &  111 \\ 
NGC67051184  &  thin  &  4282  &  1.6  &  -0.05  &  1.7  &  6.2  &  0.15  &  0.07 &  32\\ 
NGC67051423  &  thin  &  4377  &  1.8  &  0.03  &  1.9  &  7.4  &  0.19  &  0.09 &  38  \\
\hline
\hline
\end{tabular}
\tablefoot{\\ *: Infrared atlas spectrum of Arcturus \citep{Hinkle:1995} }
\end{table*}

\begin{figure*}
  \includegraphics[width=\textwidth]{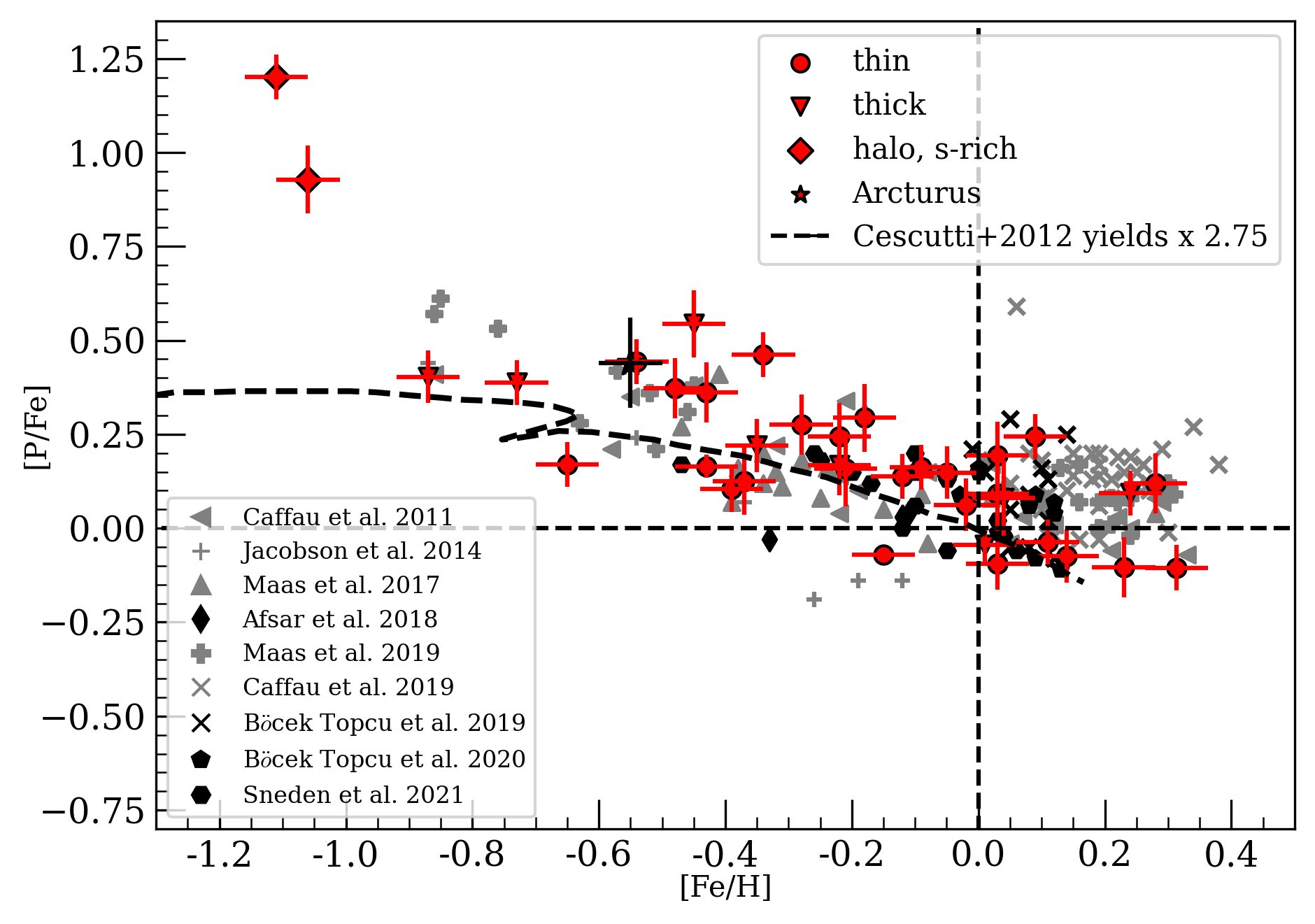}
  \caption{[P/Fe] vs [Fe/H] for stars in our sample (shown in red) with their respective stellar population represented by different symbols (thin disk - circle, thick disk - inverted triangle and halo, s-rich - diamond). Arcturus, for which P is determined from higher resolution atlas spectrum \citep{Hinkle:1995} is indicated by the black star symbol. P abundance determinations (all scaled to the solar abundance used in this work: A(P)$_{\odot}$ = 5.36 and A(Fe)$_{\odot}$ = 7.45) from multiple literature sources are also plotted in gray (giants in black) and represented by different symbols. In dashed line we show the chemical evolution trend in \cite{cescutti:2012} resulting from core collapse supernova (type II) of massive stars with the P yields from \cite{Kobayashi:2006} arbitrarily increased by a factor of 2.75   }
  \label{fig:ptrend}%
\end{figure*}

\begin{figure*}
  \includegraphics[width=\textwidth]{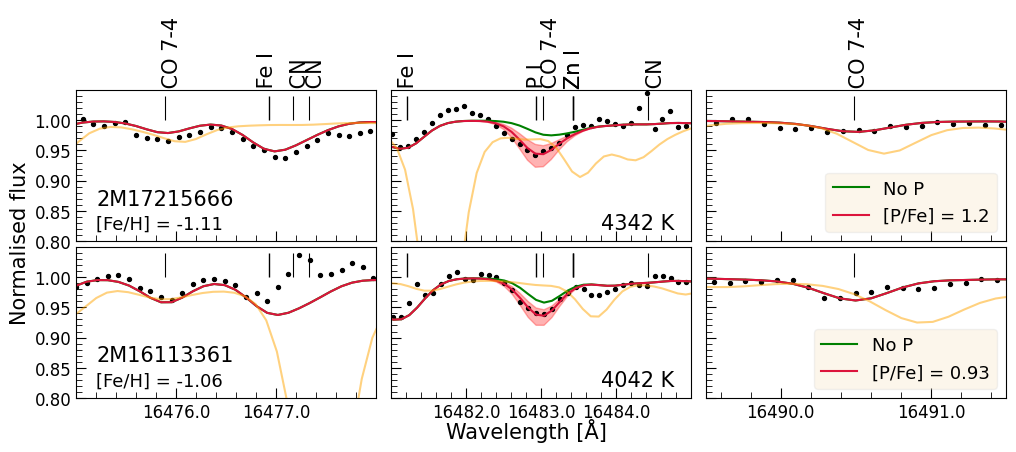}
  \caption{Same as Figure~\ref{fig:selspectra} except for the two s-rich stars in our sample.}
  \label{fig:srichspectra}%
\end{figure*}

The phosphorus abundance we determine from the line at 16482.92\,\AA\, is blended with a CO($\nu =7-4$) line. Hence it is necessary to model the CO line as well as possible to determine accurate P abundance. Fortunately, vibrational-rotational lines from the CO($\nu =7-4$) band occur close to each other at regular spacing lying at wavelengths less than 10\,\AA\, apart. We can therefore verify that the blending CO line has been reliably modeled by inspecting the neighbouring CO($\nu =7-4$) lines, which we do in the same manner as in \citet{montelius:22}.  We determine C, N, O abundances using the 25 OH ($\nu =2-0, 3-1, 4-2, 5-3$), 20 CN ($\nu =0-1, 1-2$), 10 CO lines as well as two CO band heads ($\nu =6-3, 8-5$) in our IGRINS H band spectra using SME. We find the neighbouring CO($\nu =7-4$) lines to fit well which ensures that the CO($\nu =7-4$) blend in the P feature can be properly taken into account. Table~\ref{table:all} lists the fundamental stellar parameters from optical spectra along with the C, N, O and P abundances estimated from our IGRINS spectra for the stars in our sample. We adopt the solar abundance values for phosphorus (A(P)$_{\odot}$ = 5.36) (and other elements in this study) from \cite{solar:sme} in the determination of the [P/Fe] values.

In Figure~\ref{fig:selspectra} we show the fits (red line) to the observed spectra (black points) for 4 stars chosen to represent typical spectra at metallicities of -0.87 dex (HIP63432), -0.45 dex (HIP50583), 0.01 dex (KIC3936921) and 0.28 dex (HD102328). Each row in Figure~\ref{fig:selspectra} represent one star with the lowest metallicity star in the top row and highest metallicity star in the bottom row. We show the segment of the corresponding spectrum with the P line at 16482.92\,\AA\,in the middle panel for each star (row). Synthetic and observed spectra of the neighbouring \co lines are plotted in the left and right panels to indicate the quality of the fit to the \co line species which in turn showcase the reliability of the P abundance we estimate from the \co blended P line. In addition, we show the sensitivity of the line to P abundance value with the red band that shows the variation of synthetic spectrum with $\Delta$[P/Fe] = $\pm$0.3 dex variation. This in turn gives a qualitative idea of the uncertainty on the P I feature from the synthetic spectra fit. In each panel, we plot the synthetic spectrum with no P I feature in green, i.e., showing only the CO($\nu =7-4$) line which shows the contribution of P and CO features to the line.  We also plot the telluric spectra (in orange) used for telluric correction to highlight the noise introduced during telluric correction and how our line of interest is seldomly affected by the same. Thus, Figure~\ref{fig:selspectra} shows the quality of our synthetic spectra fits and the reliability of the P abundance estimates listed in the Table~\ref{table:all}.

\begin{figure}
  \includegraphics[width=\columnwidth]{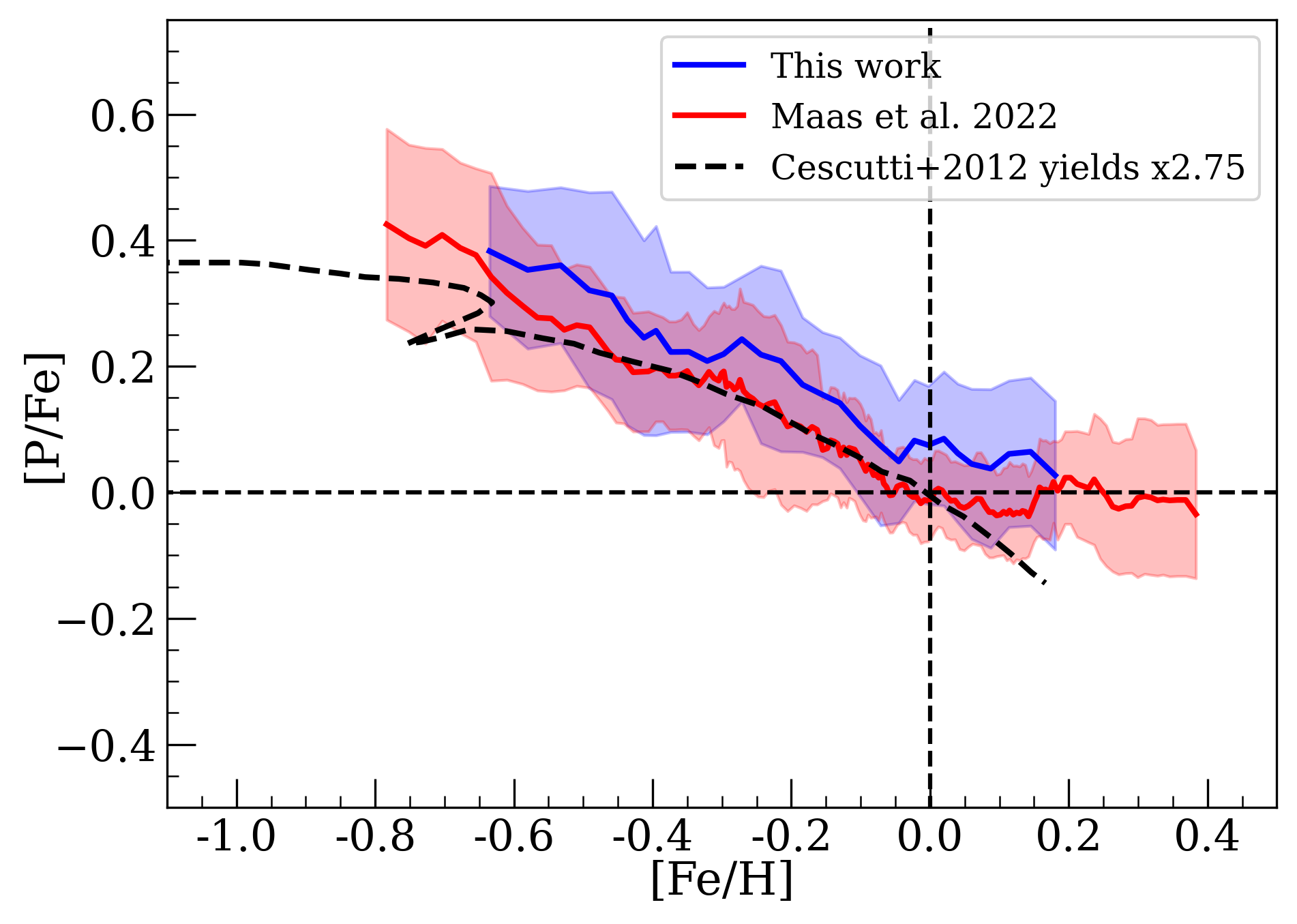}
  \caption{Running mean of [P/Fe] for the stars in our sample (blue line) and \cite{Maas:2022} sample (red line) as a function of [Fe/H]. Standard deviation is represented by the blue and red bands for our sample and \cite{Maas:2022} sample respectively. In dashed line we show the chemical evolution trend in Cescutti et al. (2012).   }
  \label{fig:ptrend_maas}%
\end{figure}

To investigate possible systematic effects such as any potential NLTE effects, we plot [P/Fe] as a function of \teff\, in Figure~\ref{fig:pteff}. We use different symbols to indicate the stellar population of each star, i.e., thin-disk, thick-disk or halo, assigned based on their separation in the [Mg/Fe] vs [Fe/H] plane \citep[see][for more details]{jonsson:17,lomaeva:19,forsberg:19}. Excluding the two halo which are s-rich and the three \teff $>$ 5000 K stars, there is no significant [P/Fe] trend with \teff. While s-rich stars have enhanced phosphorus abundances, the three 5000 K stars are metal rich and hence are expected to have lower phosphorus abundances. From the absence of any significant trend, we also conclude that the assumption that there are no significant non-LTE effects might be justified.




\subsection{Uncertainty estimates}
\label{sec:uncertain}

We determine the uncertainties in our [P/Fe] estimates that arise from the uncertainties in stellar parameters. As mentioned in the previous section, typical uncertainties in stellar parameters for our stars are estimated to be  $\pm$50K for \teff, $\pm$0.15 dex for \logg, $\pm$0.05 dex for \feh, and $\pm$0.1 km/s for $\xi_\mathrm{micro}$. Using the stellar parameter value as the mean and these typical uncertainties  as the standard deviation, we randomly generate sets of stellar parameters following a normal distribution and reanalyse the each star spectrum using those parameters. We generate 50 sets of parameters for each star. Since the CO($\nu =7-4$) lines are blended in the P feature, we redetermine the C,N,O abundances for each set of stellar parameters in order for the blending line to fit. The resulting distribution of estimated [P/Fe] values is fitted with a Gaussian function. The dispersion estimated from the resulting fit gives the uncertainty in [P/Fe], listed in the Table~\ref{table:all}. In the same manner we also determine the uncertainties in C, N, O abundances, listed in Table~\ref{table:cno}\\
We note that the sufficiently high SNR for our stars ensure that the observational noise plays a minimal role in the uncertainty while fitting the synthetic spectra. In addition, we do not find a strong correlation between SNR and [P/Fe] uncertainty for stars with similar parameters, so other factors, like carbon abundance, could play a more important role in the uncertainty.

\section{Results}
\label{sec:results}

\subsection{[P/Fe] vs [Fe/H]}
\label{sec:pfe_feh}
 

The metallicity of a star, [Fe/H], is considered to be an ideal parameter to compare with and investigate the galactic chemical evolution of phosphorus of various stellar populations. In this section, we investigate the evolution of [P/Fe] as a function of [Fe/H] and compare the resulting trend with those from the literature as well as theoretical chemical evolution models.
 
In Figure~\ref{fig:ptrend}, we plot the [P/Fe] vs [Fe/H] trend for our K giants (in red) in the metallicity range of -1.1 $\leq$ [Fe/H] $\leq$ 0.3 dex. There is a clear downward trend in [P/Fe] with increasing [Fe/H] that is slightly higher than the track of the chemical evolution model for P from \cite{cescutti:2012}. They adopted the two-infall model of galactic chemical evolution and the P massive stellar yields from \cite{Kobayashi:2006} that has been arbitrarily enhanced by a factor of 2.75 to match the observed trend in \cite{Caffau:2011}. Also this trend is similar to the trends seen for $\alpha$ elements (Si, Mg, O etc) that are produced primarily through core collapse supernovae of massive stars with short lifetimes (order of MYr). 

We estimate a [P/Fe] of 0.44 $\pm$ 0.12 dex for Arcturus from the high resolution infrared atlas spectrum of Arcturus \citep{Hinkle:1995}. We note that
from the same atlas but using the  Y band line of P I at 10581.5\AA, \cite{Maas:2017} estimated a [P/Fe] of 0.27 $\pm$ 0.1 dex (rescaled to the solar P and Fe used in our study). Furthermore,  \cite{Fanelli:2021} estimated [P/Fe] of 0.29 $\pm$ 0.06 dex (also rescaled) with a GIANO-B spectrum of Arcturus using P I lines at 10529.52 \AA\, and 10581.58 \AA. Our value is slightly higher than these, although the values overlap within uncertainties. 
 
\begin{figure*}
  \includegraphics[width=\textwidth]{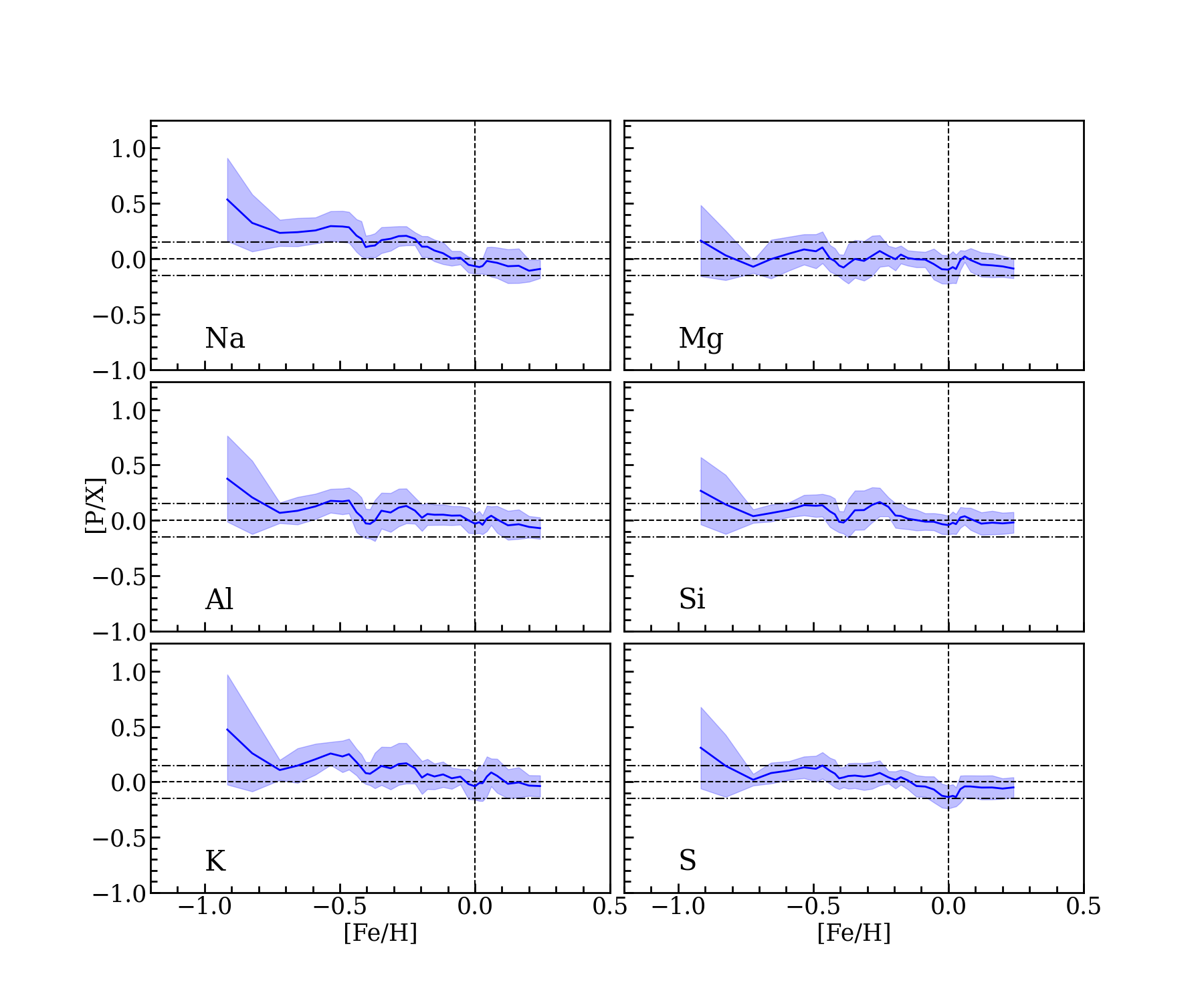}
  \caption{The running mean of the abundance ratios of P to odd-Z (Na, Al, K in in left panels) and even-Z (Mg,Si,S in right panels) as a function of [Fe/H]. The running mean and standard deviation are shown with the blue line and blue band respectively. The mean value, $<$[P/X]$>$, (excluding the two s-rich stars) is denoted in the right bottom part of each panel. Horizontal dash-dotted lines represent [P/X] values of $\pm$ 0.15 dex.}
  \label{fig:poddeven}%
 \end{figure*}
 
The [P/Fe] vs [Fe/H] trends for various samples from the literature rescaled based on the solar abundances of P and Fe used in this work are also plotted in the Figure~\ref{fig:ptrend}. We could find only a handful of studies that analyse only giants, \cite{Afsar:2018}, \cite{Topcu:2019}, \cite{Topcu:2020} and \cite{Sneden:2021}, the trends of which are shown in different color (black) compared to other works. \cite{Afsar:2018} analysed the IGRINS H- and K-band spectra of three field red horizontal branch stars with sub solar metallicity and used two P lines, 15711.5\,\AA\,and 16482.9\,\AA\, to estimate P abundance. \cite{Topcu:2019} and \cite{Topcu:2020} carried out a similar analysis of K giant stars in the open clusters NGC 6940 and NGC 752 respectively using IGRINS spectra. \cite{Sneden:2021} analysed high resolution Habitable Zone Planet Finder (HPF) spectra of 13 field red horizontal branch (RHB) stars and one open cluster giant to estimate P abundances using three lines in the 0.81-1.28 $\mu$m wavelength range. Metallicities for these stars are solar and lower, and the [P/Fe] estimates are consistent with our values.

While the APOGEE wavelength coverage include the P lines used in this work and in \cite{Afsar:2018}, the measurements of which are provided in data releases DR13 - DR16,  due to the reasons elaborated in the Section~\ref{sec:intro}, we do not include P abundances derived from the APOGEE spectra in the comparison sample.



 
 Very recently, \cite{Maas:2022} determined the P abundances for 163 stars from the P I line at 10529.52\,\AA\,based on observations with the HPF instrument on the Hobby-Eberly Telescope (HET). They find consistent abundances for both red giants and FGK dwarf stars, and $\sim$ 0.1 dex difference in [P/Fe] between thin and thick disk stars that were identified with kinematics. They also find [P/Fe] to be under-predicted by the \cite{cescutti:2012} model at lower metallicities ([Fe/H] $<$ -0.5 dex), i.e., they find higher [P/Fe] for these stars. In the Figure~\ref{fig:ptrend_maas}, we plot the running mean of the [P/Fe] as a function of [Fe/H] trends for the 163 stars from \cite{Maas:2022} (red line) and our measurements (blue line) excluding the 2 s-rich stars (see Section~\ref{sec:srich} below). The standard deviation within the 8 star bins for our sample and 15 star bins for the \cite{Maas:2022} sample are represented by the blue and red bands respectively. As mentioned above, the \cite{Maas:2022} trend closely follows the \cite{cescutti:2012} model except at lower metallicities and also at super solar metallicities. Our trend is consistently higher by $\sim$0.05 - 0.1 dex compared to both \cite{Maas:2022} and the chemical evolution model trend. Both observational trends level off at super solar metallicities as opposed to the chemical evolution trend that decreases to sub solar [P/Fe] values at these metallicities.

 
When comparing with all samples from the literature, our trend goes in unison with the trends from previous studies in the whole metallicity range within our estimated uncertainties. The [P/Fe] trend is found to be levelling off for [Fe/H] $>$ 0 in the comparison samples which is also seen for the limited number of metal rich stars in our sample. In chemical evolution models, for [P/Fe] to be constant and high at [Fe/H] $>$ 0, an additional production or adjustments to the metallicity dependence of supernovae yields may be needed.  Thus, we need to analyse more metal rich stars to confirm and understand possible mechanisms that lead to this trend.
 

 \subsection{s-rich stars}
 \label{sec:srich}

 We find a distinct enhancement of $\sim$ 0.6 - 0.9 dex for the 2 s-rich stars in our sample when compared with the \cite{cescutti:2012} model of P fitted to the \cite{Caffau:2011} P trend of solar neighborhood stars. They have been identified to be rich in s-process elements (e.g. Ba, Ce) while normal in r-process and lighter elements from the analysis of their optical spectra \citep{forsberg:19}. In Figure~\ref{fig:srichspectra}, we show the fits (red line) to the observed spectra (black points) for these 2 stars. Similar to Figure~\ref{fig:selspectra}, fits to the neighboring \co lines are shown in the left and right panels, and the red bands indicate the sensitivity of the P line to [P/Fe]. The fits along with the sensitivity band is a clear indicator that the enhancement in P abundance for these stars is real, and not due to noise or telluric residuals.
 
 \cite{Masseron:2020} identified 15 P-rich stars from APOGEE after re analysing their spectra using the BACCHUS code, finding [P/Fe] in the range of 1.2 to 2.2 dex. They also find enhanced O, Mg, Si, Al and Ce as compared to the P-normal stars. Based on further detailed analysis of the high-resolution optical spectra of two previously confirmed phosphorus-rich stars in \cite{Masseron:2020}, \cite{Masseron:2020b}  were able to find heavy-element over abundances with high first-peak (Sr, Y, and Zr) and second-peak (Ba, La, Ce, and Nd) element enhancements. While our P-rich stars are enhanced in Ce, we do not find enhancement in any other elements they report. In addition, [P/Fe] for these 2 P-rich stars are 0.93 and 1.2 dex, just below the minimum value of the [P/Fe] (1.2 dex) in the P-rich sample of \cite{Masseron:2020}. Hence, these P-rich stars might be of a different class and origin and we believe that this warrants further detailed analysis of these stars.


\section{Discussion}
\label{sec:discussion}

 \subsection{Nucleosynthesis of P}
 \label{sec:nucleosynthesis}
 
 From chemical evolution models and multi-dimensional (2D and 3D) hydro dynamical simulations of supernova explosions (Type I and II), the following formation channels for phosphorus have been suggested :
 \begin{itemize}
     \item The stable isotope of phosphorus, $^{31}$P, is considered to be mainly synthesized from neutron capture on neutron-rich Si isotopes,$^{29}$Si and $^{30}$Si, all produced in the in hydrostatic carbon and neon burning during the late stages of the evolution of massive stars (10-300 M$_{\odot}$) \citep{clayton:2003,arnett:96,Masseron:2020}. It is then released by the explosion of these massive stars  as type II supernovae with insignificant P production predicted during the explosive phases by \cite{Woosley:1995}. \citet{clayton:2003} concludes that 
     95\% of P is formed in core-collapse SN Type II.\\
     
     \item  $^{31}$P is produced, though in negligible amounts, in low mass (1-3 M$_{\odot}$) Asymptotic Giant Branch (AGB) stars by neutron capture on $^{30}$Si \citep{Karakas:2010}. This is predicted to be accompanied by large enhancements of C and s-process elements by the stellar nucleosynthesis models. 
     

 \end{itemize}
In addition, \cite{Caffau:2011} suggest phosphorus formation through proton captures on $^{30}$Si and $\alpha$ captures on $^{27}$Al. Also, \citet{weinberg:19} find a possible SNIa contribution for phosphorus in the APGOEE DR14 data, which is contrary to what chemical evolution models find \citep{cescutti:2012} with SNIa yields from \citep{iwamoto:99}. However, \citet{weinberg:19} do warn about large statistical errors for P measurements in DR14, and, as already mentioned, APOGEE decided not to include P abundances at all in the final DR17 (Holtzman et al. in prep.).

For the neutron-capture production channel for light odd-Z elements, like P, the production is sensitive to the neutron excess \citep{Burbidge:1957}, which tends to decrease with decreasing metallicity \citep{Kobayashi:2006,Caffau:2011} for metallicities lower than approximately \feh$<-1$. Our sample of stars do, unfortunately, not cover this metallicity range. 

In the sections below we attempt to decipher the nucleosynthetic origin of P based on the P abundance estimates from our sample. For this, we compare P with odd- and even-Z elements as well as investigate the primary/secondary behaviour of P by comparing A(P) with A(Mg).

 \subsection{Comparison with odd- and even-Z elements}
 \label{sec:elements}

\begin{figure}
  \includegraphics[width=\columnwidth]{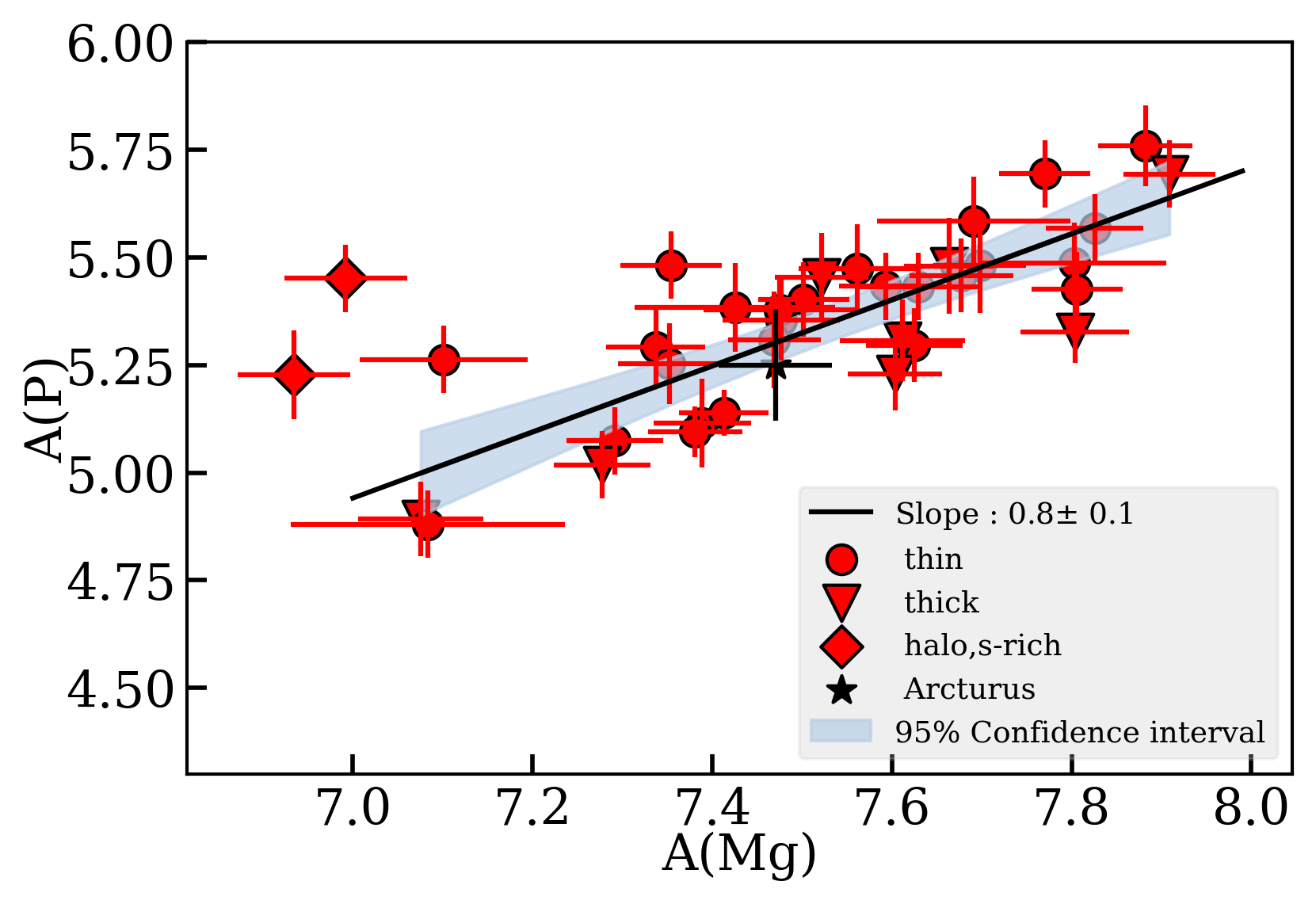}
  \caption{Abundance of phosphorus, A(P), versus abundance of magnesium, A(Mg) for stars in our sample. Excluding the 2 s-rich stars from the sample, we estimate a slope of $0.8\pm0.1$ using linear regression, with the 95\% confidence interval marked in blue. This points to a \textit{primary} behaviour of phosphorus similar to magnesium. We adopt solar abundance values of 5.36 and 7.53 for phosphorus and magnesium respectively from \cite{solar:sme}.  }
  \label{fig:apamg}%
\end{figure}

\begin{figure}
  \includegraphics[width=\columnwidth]{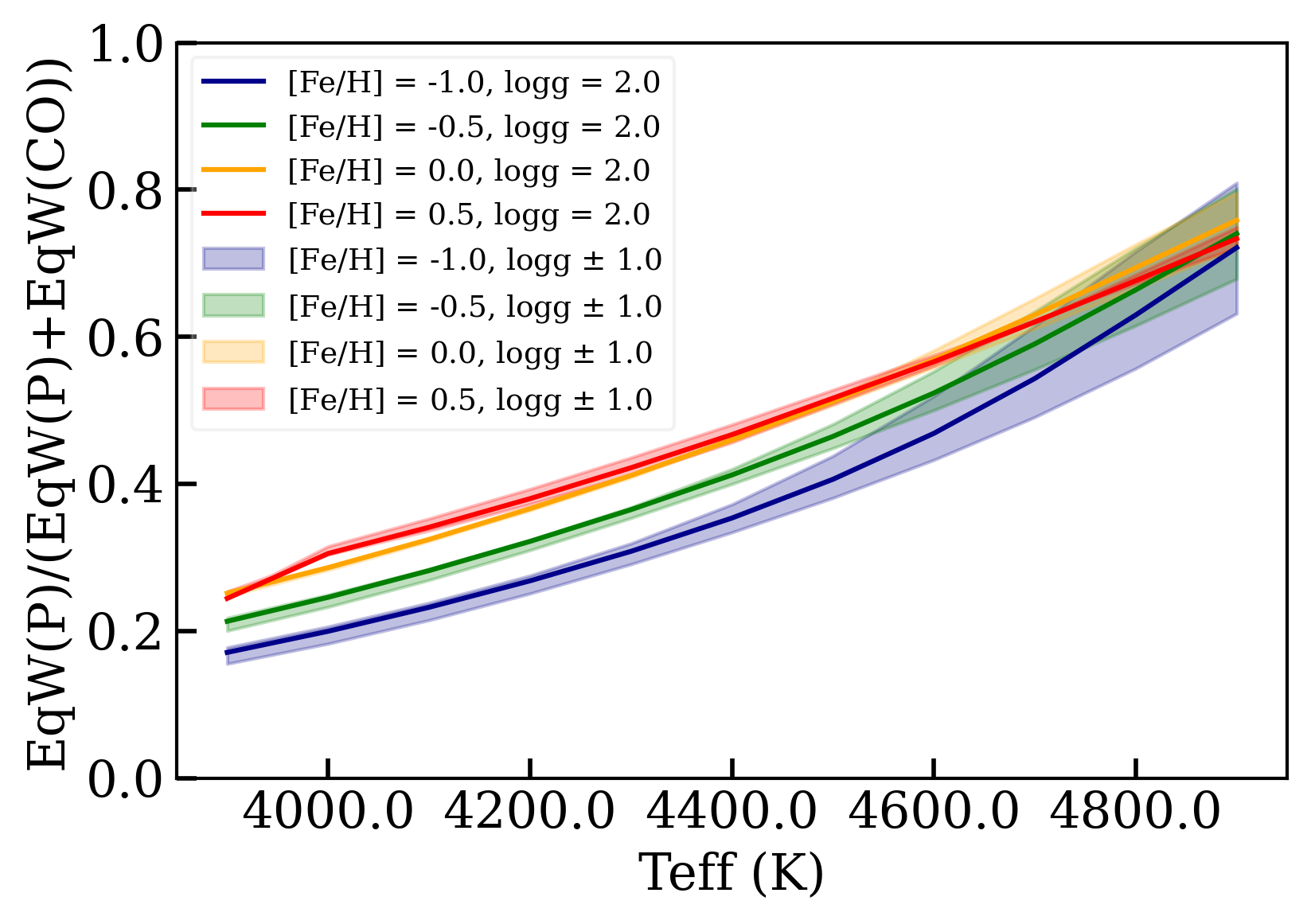}
  \caption{The fractional contribution of equivalent widths of P I line in the \co blended line versus \teff\, for a set of {\sc marcs} \citep{marcs:08} stellar models in the grid of 3900 $\leq$ \teff $\leq$ 5000 K, 1.0 $\leq$ \logg $\leq$ 3.0 and \feh of -1,-0.5, 0.0 and 0.5 dex, all with solar scaled phosphorus abundances. Different combinations of \feh\, and change in \logg\, represented by different colored lines and bands respectively with lower \logg\,  at each \feh\, resulting in lower equivalent widths for P I line.}
  \label{fig:pco}%
\end{figure}


For the comparison, we choose the odd-Z elements: Na, Al, K and the even-Z elements: Mg, Si, S, and determine their abundances from the IGRINS spectra with the lines listed in the Table~\ref{table:lines}. We carry out line-by-line analysis using SME (see Section~\ref{sec:Pabund}) and choose the mean abundance for each species by a careful visual selection of the best fitting lines. The mean abundance thus determined and the standard deviation of abundances from all chosen lines are listed in the Table~\ref{table:otherelem}. We plot the running mean of the abundance ratios of P and these elemental abundances, [P/X], as a function of [Fe/H] in Figure~\ref{fig:poddeven}, with odd-Z elements and even-Z elements in the left and right panels, respectively. In each panel, the running mean and standard deviation are shown with the blue line and blue band respectively.

We expect [P/X] ratios to be close to zero and not show any significant trends as a function of [Fe/H] if the evolution of P is similar to the element X with which it is compared. For all elements, the most metal poor bin exhibit higher difference values as the two s-enhanced stars with high P values are a major part of the stars in this bin. In the case of Mg, Al, Si, and S we find [P/X] $\sim$ 0 with the running mean values lying within [P/X] = $\pm$0.15 (horizontal dash-dot lines). The trend for [P/K] is close to 0 for [Fe/H] $>$ -0.25 dex and increase for [Fe/H] $<$ -0.25 dex. Thus [K/Fe] is not as enhanced as P in the metal poor regime. For the remaining odd-Z element, Na, the [P/X] ratio has a decreasing trend with increasing metallicity. This is because [Na/Fe] has a near constant value of zero at subsolar metallicities which starts increasing for super solar metallicities and hence show a different trend compared to P. 


 Mg, Si, and S are $\alpha$ elements that are confirmed to have been produced from core collapse supernova of massive stars \citep{Chieffi:2004,Nomoto:2013}. Similarly, Al and K are also found to have significant contribution from core collapse supernovae (for e.g. \citealt{clayton:2003,weinberg:19,Weinberg:2022}). Thus we can conclude that P behaves like the $\alpha$ elements and supports the current idea about the contribution of core collapse supernova in the origin and evolution of P. Similar investigations in previous studies have led to the same conclusion about the origin and evolution of P (see Section~\ref{sec:intro}).

 \subsection{Primary behaviour of Phosphorous for $-0.9<\feh<0.3$}
 \label{sec:Ap_Amg}
 
 Elements which are synthesized independently of the metallicity, i.e. elements produced in a star from chains of reactions starting directly from H and He are called {\it primary} elements \citep{arnett:96}. $^{16}$O, and $^{24}$Mg are examples of primary elements \citep{clayton:2003}. On the other hand, if the synthesis depends on the presence of other nuclei from earlier stellar generations, the elements produced are called {\it secondary}.
 
The synthesis of P is through neutron capture on neutron-rich Si isotopes, which are \textit{secondary} in nature \citep{clayton:2003}.  One might therefore expect phosphorus to show a {\it secondary} behaviour. In Figure \ref{fig:apamg} we therefore plot the phosphorus abundances versus the magnesium abundances. Since $^{24}$Mg, a \textit{primary} nuclei, is the main isotope of Mg (approximately 80\% \citep{clayton:2003}) we assume that our measured Mg abundance represents a primary behaviour. A \textit{secondary} behaviour for phosphorus would therefore yield a slope of 2.  In Figure \ref{fig:apamg} we, however, find a slope of $0.8\pm0.1$ for our sample using linear regression, excluding the s-element and P-rich stars, which clearly points to a \textit{primary} behaviour. We thus conclude that we find an empirically determined primary behaviour  of phosphorus at least for $-0.9<$[Fe/H]$<0.3$ range. This should be considered when  further investigating the nucleosynthetic origin of phosphorus. We note that \citet{prantzos:18} indeed find a primary behaviour in their rapidly-rotating massive stars models.




\subsection{Best stars to measure P I line at 16482.92\,\AA\,}
 \label{sec:sensitivity}
In this work, we have shown that it is possible to estimate reliable P abundances from the the P I line at 16482.92\,\AA\,even though it can be heavily blended with a \co\, line. In addition to careful manual analysis, i.e. selecting good continuum points, checking the fit to each spectrum by eye and a careful inspection of the telluric elimination procedure, we made sure to fit the neighboring \co lines well so as to determine the best P value. 

Considering the dearth of P abundance measurements in the literature, we want to investigate which type of stars should be targeted in the future to get reliable P abundance estimates from this P I line. With that aim, we estimate the equivalent widths of the P I line and \co line for a set of {\sc marcs} \citep{marcs:08} stellar models in a grid of \textit{(i)} 3900 $\leq$ \teff $\leq$ 5000 K, \textit{(ii)} 1.0 $\leq$ \logg $\leq$ 3.0 and \textit{(iii)} \feh\,  of -1.0,-0.5, 0.0 and 0.5 dex, all with solar scaled phosphorus abundances. The resulting fractional contribution of equivalent widths of P I line in the \co blended line as a function of \teff\, is shown in Figure~\ref{fig:pco} with different combinations of \feh\, and change in \logg\, represented by different colored lines and bands respectively. While the strength of P line increases with increase in \teff, the CO line strength becomes weaker at higher \teff. As is normal, metal rich giants have stronger lines and for a given metallicity, higher surface gravities, \logg, yield weaker lines. 

Thus, P lines will be stronger and the \co blend will be weaker for giant stars with \teff $>$ 4500 K and \feh $>$ -0.5 dex. For cooler metal poor stars, the contribution from the P feature is still significant enough ($>$ 15$\%$) to be measurable from a high SNR spectrum as shown in this work.

Thus we can conclude 
that the P I line at 16482.92\,\AA\, is suitable to estimate P abundances for K giants at all metallicities. Yet it has to kept in mind that the contribution from the \co blend increases with decreasing \teff\, and \feh\,. With the s-rich stars, we have also seen that this line will be useful in the hunt for more P-rich stars. 


\section{Conclusions}
\label{sec:conclusion}

Using high resolution (R$\sim$45,000) IGRINS spectra, we determine [P/Fe] for 38 K giant stars in the solar neighborhood based on a detailed spectroscopic analysis of the P I line at 16482.92\,\AA\, after a careful handling of the telluric contamination and its elimination, which we show is essential to minimize the uncertainty of the [P/Fe] trend. We show that our [P/Fe] versus [Fe/H] trend within uncertainties is consistent with the results from other studies in the literature. Our trend is higher than the chemical evolution trend from \cite{cescutti:2012} by $\sim$ 0.05 - 0.1 dex, which could mean that the enhancement factor might be higher than 2.75. We also find that the two s-rich stars in our sample exhibit clear enhancements in phosphorus abundance and warrants further detailed investigation to understand their possible origins.
We further determine abundance ratios of phosphorus with respect to other odd-Z and even-Z elements, estimated from reliable spectral lines in the H and K-band IGRINS spectra to understand the origin of phosphorus. We find that phosphorus behaves more like $\alpha$ elements than the neighboring odd-Z elements, Na and K, and thus supports the current idea about the contribution of core collapse supernova in the origin and evolution of P. In addition, we find empirical evidence that points to a primary behaviour of phosphorus even though the synthesis of P is through neutron capture on neutron-rich Si isotopes, which are secondary in nature. Finally we find that although the P I line at 16482.92\,\AA\, is suitable to estimate P abundances for K giants at all metallicities, the weak contribution of the P feature in cooler metal poor (\teff\, $<$ 4500 K, \feh\,$<$ -0.5) stars may lead to larger uncertainties in case of low SNR spectra. We also show that this line will be useful in the hunt for more P-rich stars.  
Hence, we need more observations with highly capable instruments like IGRINS that provide high resolution covering larger spectral orders in order to mitigate the dearth of reliable phosphorus abundance estimates.

\begin{acknowledgements}
We thank the anonymous referee for the very constructive comments and suggestions that improved the quality of the paper. G.N.\ acknowledges the support from the Wenner-Gren Foundations. N.R.\ acknowledge support from the Royal Physiographic Society in Lund through the Stiftelsen Walter Gyllenbergs fond and Märta och Erik Holmbergs donation. B.T.\ acknowledges the financial support from the Japan Society for the Promotion of Science as a JSPS International Research Fellow. This work used The Immersion Grating Infrared Spectrometer (IGRINS) was developed under a collaboration between the University of Texas at Austin and the Korea Astronomy and Space Science Institute (KASI) with the financial support of the US National Science Foundation under grants AST-1229522, AST-1702267 and AST-1908892, McDonald Observatory of the University of Texas at Austin, the Korean GMT Project of KASI, the Mt. Cuba Astronomical Foundation and Gemini Observatory. These results made use of the Lowell Discovery Telescope (LDT) at Lowell Observatory. Lowell is a private, non-profit institution dedicated to astrophysical research and public appreciation of astronomy and operates the LDT in partnership with Boston University, the University of Maryland, the University of Toledo, Northern Arizona University and Yale University. This paper includes data taken at The McDonald Observatory of The University of Texas at Austin.
\\
The following software and programming languages made this
research possible: TOPCAT (version 4.6; \citealt{topcat}); Python (version 3.8) and its packages ASTROPY (version 5.0; \citealt{astropy}), SCIPY \citep{scipy}, MATPLOTLIB \citep{matplotlib} and NUMPY \citep{numpy}.
\end{acknowledgements}

%
%


\bibliographystyle{aa}
\bibliography{references} 




\begin{appendix} 

\section{Tables}

\begin{table*}
\caption{ [C/Fe], [N/Fe] and [O/Fe] abundances and their uncertainties determined in this work. }\label{table:cno}
\begin{tabular}{c c c c c c c}
\hline
\hline
 Star  & [C/Fe] &  $\sigma$[C/Fe]  & [N/Fe] &  $\sigma$[N/Fe]  &  [O/Fe] &  $\sigma$[O/Fe]   \\
 \hline
  & & dex & dex & dex & dex & dex  \\
\hline
$\alpha$Boo$^{*}$  &  0.19  &  0.07  &  0.3  &  0.06  &  0.51  &  0.06\\ 
$\mu$Leo  &  0.06  &  0.05  &  0.49  &  0.05  &  0.1  &  0.05\\ 
$\epsilon$Vir  &  -0.18  &  0.09  &  0.71  &  0.04  &  0.31  &  0.04\\ 
betaGem  &  -0.08  &  0.08  &  0.38  &  0.05  &  0.07  &  0.05\\ 
HD102328  &  0.12  &  0.06  &  0.39  &  0.03  &  0.11  &  0.03\\ 
HIP50583  &  -0.03  &  0.06  &  0.39  &  0.03  &  0.19  &  0.03\\ 
HIP63432  &  0.15  &  0.07  &  0.48  &  0.06  &  0.46  &  0.06\\ 
HIP72012  &  -0.03  &  0.08  &  0.3  &  0.07  &  0.17  &  0.07\\ 
HIP90344  &  0.12  &  0.08  &  0.25  &  0.05  &  0.31  &  0.05\\ 
HIP96014  &  0.12  &  0.06  &  0.22  &  0.03  &  0.19  &  0.03\\ 
HIP102488  &  0.0  &  0.07  &  0.33  &  0.05  &  0.24  &  0.05\\ 
2M16113361  &  0.09  &  0.08  &  0.16  &  0.08  &  0.48  &  0.08\\ 
2M17215666  &  0.11  &  0.07  &  0.39  &  0.05  &  0.6  &  0.05\\ 
KIC3748585  &  -0.0  &  0.05  &  0.39  &  0.05  &  0.11  &  0.05\\ 
KIC3955590  &  0.05  &  0.03  &  0.36  &  0.05  &  0.09  &  0.05\\ 
KIC3936921  &  0.08  &  0.07  &  0.27  &  0.03  &  0.15  &  0.03\\ 
KIC4659706  &  0.14  &  0.07  &  0.34  &  0.12  &  0.17  &  0.12\\ 
KIC5113061  &  0.0  &  0.06  &  0.34  &  0.05  &  0.08  &  0.05\\ 
KIC5113910  &  0.01  &  0.06  &  0.4  &  0.05  &  0.19  &  0.05\\ 
KIC5709564  &  0.16  &  0.06  &  0.28  &  0.11  &  0.41  &  0.11\\ 
KIC5779724  &  0.17  &  0.03  &  0.28  &  0.07  &  0.43  &  0.07\\ 
KIC5859492  &  0.02  &  0.05  &  0.38  &  0.03  &  0.06  &  0.03\\ 
KIC5900096  &  0.1  &  0.07  &  0.36  &  0.05  &  0.1  &  0.05\\ 
KIC6465075  &  0.12  &  0.15  &  0.21  &  0.05  &  0.19  &  0.05\\ 
KIC6837256  &  0.21  &  0.05  &  0.3  &  0.04  &  0.48  &  0.04\\ 
KIC7006979  &  0.03  &  0.06  &  0.43  &  0.05  &  0.4  &  0.05\\
KIC10186608  &  0.11  &  0.05  &  0.21  &  0.03  &  0.16  &  0.03\\ 
KIC11045542  &  0.1  &  0.05  &  0.3  &  0.04  &  0.24  &  0.04\\ 
KIC11342694  &  0.07  &  0.05  &  0.16  &  0.03  &  0.07  &  0.03\\ 
KIC11444313  &  0.02  &  0.06  &  0.32  &  0.03  &  0.1  &  0.03\\ 
KIC11569659  &  0.16  &  0.08  &  0.26  &  0.16  &  0.29  &  0.16\\ 
$\xi$Hya  &  -0.49  &  0.05  &  0.78  &  0.04  &  0.1  &  0.04\\ 
HD175541  &  -0.08  &  0.04  &  0.38  &  0.08  &  0.24  &  0.08\\ 
HD176981  &  -0.05  &  0.05  &  0.27  &  0.08  &  0.07  &  0.08\\ 
HD206610  &  0.1  &  0.03  &  0.38  &  0.06  &  0.19  &  0.06\\ 
HD76445  &  0.08  &  0.09  &  0.37  &  0.03  &  0.32  &  0.03\\ 
NGC67051184  &  0.02  &  0.07  &  0.32  &  0.04  &  0.03  &  0.04\\ 
NGC67051423  &  0.01  &  0.07  &  0.35  &  0.04  &  0.01  &  0.04\\ 
\hline
\hline
\end{tabular}
\end{table*}

\begin{table}
  \begin{center}
\caption{Wavelengths of the lines for the elements Mg, Al, Si, S and K in IGRINS H and K-bands used to determined their respective abundances for the stars in our sample. }\label{table:lines}
\begin{tabular}{l c c c c c c c}
 Element & Wavelength (\AA\,)  \\
\hline
\hline
\multirow{7}{*}{Na} & 16373.87  \\
 &  16388.85   \\
 &  17227.70   \\
 &  22056.43   \\
 &  22083.66   \\
 &  23348.42   \\
 &  23379.14   \\
\hline
\hline
\multirow{3}{*}{Mg} & 15740.70    \\
 &  15748.89    \\
 &  17108.63    \\ 
\hline
\hline
\multirow{4}{*}{Al} & 16718.97      \\
 &  16750.60    \\
 &  16763.37    \\ 
 &  17699.05    \\ 
\hline
\hline
\multirow{5}{*}{Si} & 15361.16       \\
 &  16163.71     \\
 &  16343.93     \\ 
 &  16828.18     \\ 
 &  17225.63     \\
\hline
\hline
\multirow{9}{*}{S} & 15469.82     \\
 &  15478.48     \\
 &  22006.41    \\
 &  22507.60     \\
 &  22519.11     \\
 &  22526.05    \\
 &  22552.61    \\
 &  22563.87      \\
 &  22575.43    \\
\hline
\hline
\multirow{2}{*}{K} & 15163.09       \\
 &  15168.40    \\
\hline
\hline
\end{tabular}
\end{center}
\end{table}

 \begin{table*}
\begin{center}
\caption{ Mean abundances of Na, Mg, Al, Si, S and K and their uncertainties based on line-by-line analysis using the lines listed in the Table~\ref{table:lines}  } \label{table:otherelem}
\begin{tabular}{c c  c  c  c  c  c c }
\hline
    &   &  &   [X/Fe] $\pm$ $\sigma$[X/Fe]  &   &  &      \\
\hline
  Star  & Na   & Mg    &  Al   & Si   & S  & K   \\
 \hline
 $\alpha$Boo  &  0.1  $\pm$  0.08  &  0.49  $\pm$  0.04  &  0.29  $\pm$  0.07  &  0.3 $\pm$ 0.07  &  0.37 $\pm$ 0.19  &  0.27 $\pm$ 0.02\\ 
$\mu$Leo  &  0.24 $\pm$ 0.11  &  0.16 $\pm$ 0.03  &  0.1 $\pm$ 0.05  &  0.0 $\pm$ 0.09  &  0.1 $\pm$ 0.11  &  0.08 $\pm$ 0.08\\ 
EpsVir  &  0.19 $\pm$ 0.06  &  -0.05 $\pm$ 0.01  &  0.0 $\pm$ 0.03  &  0.03 $\pm$ 0.01  &  0.06 $\pm$ 0.07  &  0.04 $\pm$ 0.02\\ 
$\beta$Gem  &  0.09 $\pm$ 0.06  &  -0.01 $\pm$ 0.01  &  -0.03 $\pm$ 0.05  &  -0.01 $\pm$ 0.04  &  0.0 $\pm$ 0.11  &  0.04 $\pm$ 0.05\\ 
HD102328  &  0.16 $\pm$ 0.08  &  0.07 $\pm$ 0.02  &  0.06 $\pm$ 0.04  &  0.0 $\pm$ 0.07  &  0.04 $\pm$ 0.07  &  0.04 $\pm$ 0.09\\ 
HIP50583  &  0.02 $\pm$ 0.08  &  0.11 $\pm$ 0.08  &  0.14 $\pm$ 0.08  &  0.28 $\pm$ 0.05  &  0.19 $\pm$ 0.19  &  0.01 $\pm$ 0.11\\ 
HIP63432  &  0.02 $\pm$ 0.04  &  0.42 $\pm$ 0.05  &  0.2 $\pm$ 0.12  &  0.38 $\pm$ 0.04  &  0.45 $\pm$ 0.24  &  0.17 $\pm$ 0.03\\ 
HIP72012  &  -0.01 $\pm$ 0.04  &  0.23 $\pm$ 0.04  &  0.06 $\pm$ 0.05  &  0.03 $\pm$ 0.06  &  0.24 $\pm$ 0.16  &  -0.0 $\pm$ 0.0\\ 
HIP90344  &  0.04 $\pm$ 0.05  &  0.15 $\pm$ 0.02  &  0.13 $\pm$ 0.08  &  0.11 $\pm$ 0.06  &  0.12 $\pm$ 0.1  &  0.08 $\pm$ 0.08\\ 
HIP96014  &  0.05 $\pm$ 0.05  &  0.24 $\pm$ 0.02  &  0.17 $\pm$ 0.04  &  0.2 $\pm$ 0.03  &  0.14 $\pm$ 0.12  &  0.09 $\pm$ 0.03\\ 
HIP102488  &  0.04 $\pm$ 0.05  &  0.15 $\pm$ 0.01  &  0.14 $\pm$ 0.04  &  0.11 $\pm$ 0.01  &  0.07 $\pm$ 0.07  &  0.16 $\pm$ 0.03\\ 
2M16113361  &  0.13 $\pm$ 0.23  &  0.46 $\pm$ 0.04  &  0.08 $\pm$ 0.14  &  0.25 $\pm$ 0.17  &  0.23 $\pm$ 0.4  &  0.0 $\pm$ 0.0\\ 
2M17215666  &  0.07 $\pm$ 0.19  &  0.57 $\pm$ 0.05  &  0.36 $\pm$ 0.21  &  0.6 $\pm$ 0.03  &  0.39 $\pm$ 0.31  &  0.0 $\pm$ 0.0\\ 
KIC3748585  &  0.1 $\pm$ 0.08  &  0.06 $\pm$ 0.02  &  0.06 $\pm$ 0.03  &  0.03 $\pm$ 0.04  &  0.12 $\pm$ 0.1  &  0.16 $\pm$ 0.04\\ 
KIC3936921  &  0.0 $\pm$ 0.07  &  0.26 $\pm$ 0.03  &  0.06 $\pm$ 0.02  &  0.11 $\pm$ 0.03  &  0.18 $\pm$ 0.12  &  0.03 $\pm$ 0.1\\ 
KIC3955590  &  0.14 $\pm$ 0.08  &  0.14 $\pm$ 0.01  &  0.1 $\pm$ 0.05  &  0.09 $\pm$ 0.06  &  0.21 $\pm$ 0.15  &  0.07 $\pm$ 0.04\\ 
KIC4659706  &  0.06 $\pm$ 0.08  &  0.14 $\pm$ 0.01  &  0.07 $\pm$ 0.07  &  0.04 $\pm$ 0.04  &  0.08 $\pm$ 0.1  &  0.02 $\pm$ 0.05\\ 
KIC5113061  &  0.16 $\pm$ 0.07  &  0.19 $\pm$ 0.04  &  0.01 $\pm$ 0.04  &  0.12 $\pm$ 0.05  &  0.24 $\pm$ 0.19  &  0.04 $\pm$ 0.04\\ 
KIC5113910  &  0.17 $\pm$ 0.04  &  0.3 $\pm$ 0.03  &  0.2 $\pm$ 0.05  &  0.25 $\pm$ 0.04  &  0.26 $\pm$ 0.17  &  0.1 $\pm$ 0.1\\ 
KIC5709564  &  0.12 $\pm$ 0.09  &  0.42 $\pm$ 0.02  &  0.28 $\pm$ 0.04  &  0.3 $\pm$ 0.04  &  0.25 $\pm$ 0.16  &  0.25 $\pm$ 0.0\\ 
KIC5779724  &  0.12 $\pm$ 0.09  &  0.44 $\pm$ 0.02  &  0.28 $\pm$ 0.06  &  0.27 $\pm$ 0.04  &  0.31 $\pm$ 0.19  &  0.27 $\pm$ 0.04\\ 
KIC5859492  &  0.06 $\pm$ 0.04  &  0.15 $\pm$ 0.01  &  0.06 $\pm$ 0.06  &  0.09 $\pm$ 0.03  &  0.12 $\pm$ 0.11  &  0.02 $\pm$ 0.02\\ 
KIC5900096  &  0.1 $\pm$ 0.08  &  0.04 $\pm$ 0.09  &  0.07 $\pm$ 0.05  &  0.02 $\pm$ 0.04  &  0.06 $\pm$ 0.08  &  0.01 $\pm$ 0.03\\ 
KIC6465075  &  0.07 $\pm$ 0.07  &  0.23 $\pm$ 0.02  &  0.17 $\pm$ 0.03  &  0.19 $\pm$ 0.03  &  0.11 $\pm$ 0.15  &  0.09 $\pm$ 0.09\\ 
KIC6837256  &  0.16 $\pm$ 0.11  &  0.48 $\pm$ 0.02  &  0.38 $\pm$ 0.02  &  0.33 $\pm$ 0.05  &  0.3 $\pm$ 0.16  &  0.37 $\pm$ 0.03\\ 
KIC7006979  &  0.13 $\pm$ 0.14  &  0.16 $\pm$ 0.03  &  0.12 $\pm$ 0.07  &  0.1 $\pm$ 0.03  &  0.18 $\pm$ 0.26  &  0.04 $\pm$ 0.04\\ 
KIC10186608  &  0.01 $\pm$ 0.07  &  0.06 $\pm$ 0.07  &  0.08 $\pm$ 0.03  &  0.16 $\pm$ 0.06  &  0.07 $\pm$ 0.11  &  0.06 $\pm$ 0.07\\ 
KIC11045542  &  0.09 $\pm$ 0.13  &  0.2 $\pm$ 0.14  &  0.18 $\pm$ 0.11  &  0.19 $\pm$ 0.09  &  0.18 $\pm$ 0.21  &  0.04 $\pm$ 0.04\\ 
KIC11342694  &  0.04 $\pm$ 0.12  &  0.14 $\pm$ 0.01  &  0.1 $\pm$ 0.03  &  0.05 $\pm$ 0.04  &  0.06 $\pm$ 0.14  &  0.06 $\pm$ 0.06\\ 
KIC11444313  &  0.09 $\pm$ 0.05  &  0.21 $\pm$ 0.04  &  0.16 $\pm$ 0.05  &  0.15 $\pm$ 0.05  &  0.21 $\pm$ 0.14  &  0.1 $\pm$ 0.01\\ 
KIC11569659  &  0.11 $\pm$ 0.1  &  0.28 $\pm$ 0.01  &  0.2 $\pm$ 0.05  &  0.21 $\pm$ 0.05  &  0.23 $\pm$ 0.2  &  0.17 $\pm$ 0.06\\ 
$\xi$Hya  &  0.05 $\pm$ 0.08  &  -0.02 $\pm$ 0.02  &  -0.02 $\pm$ 0.05  &  -0.08 $\pm$ 0.1  &  -0.07 $\pm$ 0.14  &  -0.01 $\pm$ 0.02\\ 
HD175541  &  0.0 $\pm$ 0.0  &  0.03 $\pm$ 0.0  &  0.04 $\pm$ 0.04  &  0.07 $\pm$ 0.04  &  0.02 $\pm$ 0.02  &  0.08 $\pm$ 0.01\\ 
HD176981  &  0.07 $\pm$ 0.09  &  0.12 $\pm$ 0.1  &  0.06 $\pm$ 0.04  &  0.08 $\pm$ 0.08  &  0.18 $\pm$ 0.03  &  0.02 $\pm$ 0.0\\ 
HD206610  &  0.06 $\pm$ 0.05  &  0.09 $\pm$ 0.0  &  0.11 $\pm$ 0.03  &  0.07 $\pm$ 0.01  &  0.07 $\pm$ 0.01  &  0.07 $\pm$ 0.01\\ 
HD76445  &  0.04 $\pm$ 0.05  &  0.3 $\pm$ 0.05  &  0.27 $\pm$ 0.02  &  0.16 $\pm$ 0.09  &  0.23 $\pm$ 0.01  &  0.25 $\pm$ 0.02\\ 
NGC67051184  &  0.15 $\pm$ 0.09  &  0.2 $\pm$ 0.03  &  0.12 $\pm$ 0.07  &  0.16 $\pm$ 0.05  &  0.32 $\pm$ 0.0  &  0.05 $\pm$ 0.02\\ 
NGC67051423  &  0.26 $\pm$ 0.09  &  0.13 $\pm$ 0.09  &  0.1 $\pm$ 0.04  &  0.1 $\pm$ 0.06  &  0.33 $\pm$ 0.04  &  -0.05 $\pm$ 0.01\\ 
\hline
\hline
\end{tabular}
\end{center}
\end{table*}

\end{appendix}

\end{document}